\documentclass[preprint,superscriptaddress,showpacs,preprintnumbers,amsmath,amssymb]{revtex4}

\usepackage{graphicx}
\usepackage{dcolumn}
\usepackage{bm}

\begin{document}


\title{Landau damping of Bogoliubov excitations in optical lattices at finite temperature}

\author{Shunji Tsuchiya}
 \email{stuchiya@iis.u-tokyo.ac.jp}
\affiliation{Department of Physics, University of Toronto, Toronto, Ontario, Canada M5S 1A7}
\affiliation{Department of Physics, Waseda University, 3-4-1 Okubo, Tokyo 169-8555, Japan.}
\altaffiliation[Present address ]{Institute of Industrial Science, University of Tokyo, 4-6-1 Komaba, Megura-ku, Tokyo 153-8505, Japan.}
\author{Allan Griffin}
\email{griffin@physics.utoronto.ca}
\affiliation{Department of Physics, University of Toronto, Toronto, Ontario, Canada M5S 1A7}%

\date{\today}

\begin{abstract}
We study the damping of Bogoliubov excitations in an optical lattice at finite temperatures.
For simplicity, we consider a Bose-Hubbard tight-binding model and limit our analysis to the lowest excitation band.
We use the Popov approximation to calculate the temperature dependence of the number of condensate atoms $n^{\mathrm c 0}(T)$ in each lattice well.
We calculate the Landau damping of a Bogoliubov excitation in an optical lattice due to coupling to a thermal cloud of excitations. While most of the paper concentrates on 1D optical lattices, we also briefly present results for 2D and 3D lattices.
For energy conservation to be satisfied, we find that the excitations 
in the collision process must exhibit anomalous dispersion ({\it i.e.} the excitation energy must bend upward at low momentum),
as also exhibited by phonons in superfluid $^4\rm{He}$.  
This leads to the sudden disappearance of all damping processes 
in $D$-dimensional simple cubic optical lattice
when $U n^{\rm c 0}\ge 6DJ$, where $U$ is the on-site interaction, 
and $J$ is the hopping matrix element. Beliaev damping in a 1D optical lattice is briefly discussed. 
\end{abstract}

\maketitle

\section{\label{intro}Introduction}

Bose-Einstein condensates in a periodic optical lattice potential 
have attracted much interest. 
An optical lattice is an ideal system for studying a variety of solid state physics phenomena, such as Bloch oscillations \cite{Morsch,Roati}, Landau-Zener tunneling \cite{Anderson, Jona-Lasinio}, and Josephson oscillations \cite{Inguscio1}. An optical lattice is an ideal crystal without lattice imperfections or impurities, and parameters can be changed easily over a wide range.
The experimental achievement of the superfluid-Mott insulator 
transition in an optical lattice \cite{Greiner} also proved 
the usefulness of optical lattices to understand basic properties of strongly correlated systems. 
Moreover, Bose condensates in an optical lattice are predicted to 
show novel phenomena such as dynamical instability \cite{Wu2,Smerzi3,TaylorZaremba, Fallani, Modugno} 
and swallow-tail energy loops in the band structure \cite{Wu1,Diakonov,Machholm}.
Recent studies have concentrated on the case of pure Bose condensates at zero temperature.
The effect of a thermal cloud on the dynamics of the condensate excitations  
in an optical lattice have not been studied very much. This is the subject of the present paper.

Recently, several experimental papers have reported results on the collective modes of Bose condensates in a one-dimensional optical lattice \cite{Inguscio3, Inguscio2,Esslinger2,Porto} and its damping at finite temperature \cite{Inguscio2}.
In these experiments, a shift of the oscillation frequency in the presence of 
the optical lattice and a sharp change of the damping rate with increasing depth of the optical lattice have been observed.
The measured frequency shift \cite{Inguscio3} of the collective modes is in good agreement with the renormalized mass theory of Kr\"amer {\it et al.} \cite{Kramer1}. The damping of condensate oscillations in a 1D optical lattice at $T=0$ has been measured \cite{Esslinger2}. The damping of the condensate excitations at finite $T$ has not been studied in any detail.

Since our major interest in this paper is the thermal cloud of excitations, 
it is important to make a clear distinction at the outset between (a) The Bloch-Bogoliubov excitations associated with linearized fluctuations of an equilibrium Bose-condensate 
and (b) The stationary states of the time-independent Gross-Pitaevskii equation for the Bose order parameter.
In a continuum model, the latter states can be described by the eigenfunction
(we use a 1D lattice for illustration)
\begin{eqnarray}
\Phi_k^0(x)=e^{\mathrm{i}kx}u_k(x),\label{Blochcondensate}
\end{eqnarray}
where the condensate Bloch function satisfies the usual periodicity condition $u_k(x)=u_k(x+ld)$, where $d$ is the optical lattice spacing and $l$ is an integer.
Physically, $\Phi_k^0(x)$ corresponds to a solution of the GP equation with a superfluid flow in the periodic potential with the condensate quasi-momentum $k$.
Recent theoretical literature (see, for example, Refs. \cite{Wu2} and \cite{TaylorZaremba}) has reported extensive studies of such condensate Bloch states, including 
their energy band structure and stability.
The latter question can be studied by considering the dynamic fluctuations $\delta \Phi_k(x,q)$ around the equilibrium state $\Phi_k^0(x)$, 
with the generalized Bogoliubov excitation energy $E_q(k)$.
These excitations are also described by a quasi-momentum $q$ in the first Brillouin zone (BZ) and will be referred to as the Bloch-Bogoliubov excitations of the optical lattice.
The thermal cloud of non-condensate atoms which is present at finite temperature is described as a gas of these Bloch-Bogoliubov excitations.
As emphasized in the literature \cite{Pitaevskii&Stringari}, one must not confuse the energy bands of these Bloch-Bogoliubov excitations $E_q(k)$ with the condensate energy bands or Bloch eigenstates described by Eq. (\ref{Blochcondensate}).
That is, we must distinguish between the ``condensate" energy band and the ``excitation" energy band.
In our tight-binding model, the analogue of Eq. (\ref{Blochcondensate}) is 
\begin{eqnarray}
\Phi_k^0(l)=e^{\mathrm{i}kld}\sqrt{n^{\mathrm c}(k)},\label{latticeBlochcondensate}
\end{eqnarray}
where $d$ is the optical lattice spacing and $l$ is an integer.
While one could generalize our analysis, we only consider the Bose condensate in the $k=0$ Bloch state, in which case Eq. (\ref{latticeBlochcondensate}) reduces to
\begin{eqnarray}
\Phi_{k=0}^0(l)=\sqrt{n^{\mathrm c 0}}.
\end{eqnarray}
Here $n^{\mathrm c 0}$ denotes the number of condensate atoms trapped in each well of the optical lattice, labelled by $l$.

In this paper, we study the dynamics of Bose atoms in an optical lattice 
at finite temperatures.
We discuss a Bose-Hubbard tight-binding model using the Gross-Pitaevskii approximation, but generalized to include the presence of non-condensate atoms. 
We calculate the temperature dependence of the condensate atom number 
$n^{\rm c 0}(T)$ in each lattice well using the static Popov approximation.
This is needed in our calculation of the Landau damping of condensate modes due to coupling to thermal excitations.
We also extend our earlier results for a 1D optical lattice \cite{TsuchiyaGriffin} 
to 2D and 3D optical lattices.
As discussed in Ref. \cite{TsuchiyaGriffin}, for damping processes to occur, the dispersion relation of 
the Bloch-Bogoliubov excitations $E_q$ must initially bend upward 
as the quasi-momentum $q$ increases. 
This is referred to as ``anomalous dispersion'' and 
is also the source of 3-phonon damping of long wavelength phonons 
in superfluid $^4\rm{He}$ \cite{anomalous1,anomalous2}.
This condition leads to a dramatic disappearance of all damping processes of 
phonon modes in a $D$-dimensional optical lattice when $\alpha\equiv Un^{\rm c 0}/J >6D$, where $U$ is the on-site interaction and $J$ is the hopping matrix element.

This paper is organized as follows. 
In Sec. \ref{model} we introduce the well-known Bose-Hubbard tight-binding model which describes Bose gases in an optical lattice.
In Sec. \ref{generalizedGP}, we discuss the generalized discrete Gross-Pitaevskii equation 
and the characteristic changes in the Bloch-Bogoliubov excitation dispersion relation as a function of the dimensionless interaction parameter $\alpha=Un^{\rm c 0}/J$.
In Sec. \ref{staticPopov}, we introduce the static Popov approximation for optical lattices, and calculate the condensate fraction and $\alpha$ as a function of the optical lattice depth and temperature.
In Sec. \ref{damping}, we calculate the Landau damping of Bogoliubov excitations in an optical lattice of dimension $D$. We also briefly remark on Beliaev damping into two excitations.
We made some concluding remarks in Sec. \ref{conclusion}

As mentioned, some of results of this paper in a 1D optical lattice were briefly described in Ref. \cite{TsuchiyaGriffin}. In the present paper, we give a more detailed derivation and discussion, as well as new results for 2D and 3D optical lattices. 

\section{\label{model}model}

We consider bosonic atoms in an optical lattice potential
\begin{eqnarray}
V_\mathrm{op}(\mathbf{r})=sE_R\sum_{i=1}^{D}\sin^2(k x_i), 
\end{eqnarray}
where $s$ is the usual dimensionless parameter describing the optical lattice depth in units of the photon recoil energy $E_R\equiv \hbar^2 k^2/2m$. 
$D$ is the dimension of the optical lattice and $d=\frac{\pi}{k}=\frac{\lambda}{2}$ is the lattice period.
We only consider simple cubic lattices considered in recent experiments \cite{Esslinger1,Greiner}.
We call attention to the recent technique \cite{Esslinger2} of producing a two-dimensional array of long, tightly confined condensate tubes by loading a Bose condensate into a deep 2D optical lattice potential, which prevents atoms from hopping between different tubes. 
With an additional 1D optical lattice potential along a tube, an ideal 1D system can be experimentally realized \cite{Esslinger2}. One can also have an ideal 2D system by loading a condensate into a deep 1D optical lattice and a shallow 2D optical lattice.
We assume this experimental setup for the realization of 1D and 2D optical lattices in the present paper.  
We also assume that the laser intensity determining the depth of the optical 
lattice wells is large enough to make the atomic wave functions well localized on the individual sites ({\it i.e.}, where we can use the tight-binding approximation). 
The energy gap between the first and the second excitation bands is large compared to the thermal energy ($2k_\mathrm{B}T/E_R\ll s$), and thus only the first band is thermally occupied.

Within a tight-binding approximation, the Hamiltonian is effectively 
described by the Bose-Hubbard model \cite{Fisher,Jaksch,Stoof,ReyBurnett} as 
\begin{eqnarray}
H=-J\sum_{<j,l>}(a_{j}^\dagger a_{l}^{}
 + a_{l}^\dagger a_{j}^{}) +\frac{1}{2}U\sum_{j} a_{j}^\dagger a_{j}^\dagger
a_{j}^{} a_{j}^{},\label{H}
\end{eqnarray}
where $a_j$ and $a_j^\dagger$ are destruction and creation operators of atoms
on the $j$-th lattice site.
$\langle j,l \rangle$ represents nearest neighbor pairs of lattice sites.
The first term describes the kinetic energy due to the hopping of atoms between sites. 
The hopping matrix element $J$ is given by 
\begin{eqnarray}
J=-\int d\mathbf{r} w_j^\ast(\mathbf{r}) \left( -\frac{\hbar^2\nabla^2}{2m}+V_\mathrm{op}(\mathbf{r})\right)w_{l}(\mathbf{r}),
\end{eqnarray}
where $w_j(\mathbf{r})$ is a wave function localized on the $j$-th lattice site, and $m$ is the atomic mass. 
Expanding the optical lattice potential around the minima of the potential wells, the well trap frequency is $\omega_s\equiv s^{1/2}\frac{\hbar k^2}{m}$. 
Approximating the localized function as the ground state wave function of a harmonic oscillator with frequency $\omega_s$ at the potential minima of $j$-th site 
\begin{eqnarray}
w_j(\mathbf{r})=\left(\frac{m\omega_s}{\pi\hbar}\right)^{D/4}\exp\left(-\frac{m\omega_s}{2\hbar}(\mathbf{r}-\mathbf{r}_j)^2\right),
\end{eqnarray}
one obtains
\begin{eqnarray}
\frac{J}{E_R}\sim\left[(\frac{\pi^2 s}{4}-\frac{s^{1/2}}{2})-\frac{1}{2}s(1+\exp(-s^{-1/2}))\right] e^{-\pi^2 s^{1/2}/4}.\label{hopping}
\end{eqnarray}
Here $s^{1/2}\sim \frac{\sqrt{2m(sE_R)}}{\hbar}d$ can be interpreted as a WKB factor for tunneling in an optical lattice potential which has height $sE_R$ and well width $d$.

The second term in Eq. (\ref{H}) describes the interaction between atoms 
when they are at the same site. 
We assume that atoms can move along $z$ direction in 1D case and in $xy$ plane in 2D case.
The on-site interaction $U$ depends on the dimensionality of the optical lattice. $U$ is given by \cite{Jaksch}
\begin{eqnarray}
U=g \int \mathrm d\mathbf{r}_\bot \left|\phi_\bot(\mathbf{r}_\bot)\right|^4\int \mathrm d{z}|w_j(z)|^4, \hspace{1cm} &(\mathrm{1D})&\\
=g \int \mathrm dz \left|\phi_\|(z)\right|^4\int \mathrm d x \mathrm d y|w_j(x,y)|^4, \hspace{0.6cm}&(\mathrm{2D})&\\
=g\int \mathrm d\mathbf{r}|w_j(\mathbf{r})|^4,  \hspace{3.9cm}&(\mathrm{3D})&
\end{eqnarray}
where $g=4\pi\hbar^2a/m$ and $a$ is the $s$-wave scattering length. 
Here $\phi_\bot(\mathbf{r}_\bot)=\left(\frac{m\omega_\bot}{\pi\hbar}\right)^{1/2}\exp\left(-\frac{m\omega_\bot}{2\hbar}\mathbf{r}_\bot^2\right)$ and $\phi_\|(z)=\left(\frac{m\omega_\|}{\pi\hbar}\right)^{1/4}\exp\left(-\frac{m\omega_\|}{2\hbar}z^2\right)$ are the ground state wave functions in optical lattice well traps for confining atoms in 1 and 2 dimensions.
Approximating the localized function $w_j$ as a simple Gaussian, one obtains \cite{Jaksch}
\begin{eqnarray}
\frac{U}{E_R}\sim\frac{g}{(2\pi)^{3/2}a_\bot^2 a_s}
=\frac{2^{3/2}a d}{\pi^{3/2}a_\bot^2}s^{1/4},\hspace{1cm}&(\mathrm{1D})&\\
\sim\frac{g}{(2\pi)^{3/2}a_\| a_s^2}
=\frac{2^{3/2}a}{\pi^{1/2}a_\|}s^{1/2},\hspace{1cm}&(\mathrm{2D})& \\
\sim\frac{g}{(2\pi)^{3/2}a_s^3}
=\frac{2^{3/2}\pi^{1/2}a}{d}s^{3/4},\hspace{1cm}&(\mathrm{3D})&
\end{eqnarray} 
where $a_s=\sqrt{\frac{\hbar}{m\omega_s}}$, $a_\bot=\sqrt{\frac{\hbar}{m\omega_\bot}}$, and $a_\|=\sqrt{\frac{\hbar}{m\omega_\|}}$. 

\section{\label{generalizedGP}the generalized Gross-Pitaevskii equation}

In order to investigate the collective excitations of a Bose condensate 
in an optical lattice, we restrict ourselves to the superfluid phase of the system described by Eq. (\ref{H}). 
We start from the Heisenberg equation of motion for $a_j(t)$ (we set $\hbar=1$ from now on),
\begin{eqnarray}
\mathrm{i}\frac{\mathrm d}{\mathrm d t}a_j(t)&=&[a_j(t),H]\\
&=&-J\sum_{\langle l \rangle}a_{l}(t)+Ua_j^\dagger(t) a_j(t) a_j(t),
\end{eqnarray}
where $\langle l \rangle$ means that $l$ runs over the nearest neighbor sites of site $j$.
In the presence of Bose condensation, we write $a_j(t)=\Phi_j(t)+\tilde{\psi}_j(t)$, 
where $\Phi_j(t)=\langle a_j(t)\rangle$ is the condensate wave function at site $j$ and 
$\tilde{\psi}_j(t)$ is the non-condensate field operator (with $\langle\tilde{\psi}_j(t) \rangle=0$).
This non-condensate operator $\tilde{\psi}_j$ satisfies the usual Bose commutation relations,
\begin{eqnarray}
[\tilde{\psi}_j,\tilde{\psi}_l^\dagger]=\delta_{jl}, \ \ [\tilde{\psi}_j,\tilde{\psi}_l]=[\tilde{\psi}_j^\dagger,\tilde{\psi}_l^\dagger]=0.
\end{eqnarray}
The equation for the condensate wave function is obtained by taking
an average of the Heisenberg equation.
If, as usual, we neglect the anomalous correlations $m_j(t)=\langle\tilde{\psi}_j \tilde{\psi}_j\rangle$ and $\langle\tilde{\psi}_j^\dagger \tilde{\psi}_j\tilde{\psi}_j\rangle$,
we are left with a ``generalized" form of the discrete Gross-Pitaevskii equation \cite{Hutchinson} which includes the mean field of non-condensate atoms
\begin{eqnarray}
\mathrm{i}\frac{\mathrm d}{\mathrm d t}\Phi_j(t)=-J\sum_{\langle l \rangle}\Phi_{l}(t)
+U(n_j^\mathrm{c}(t)+2\tilde{n}_j(t))\Phi_j(t).
\label{GP}
\end{eqnarray}
Here, $n_j^\mathrm{c}(t)$ is the number of condensate atoms on the $j$-th lattice site and $\tilde{n}_j(t)=\langle\tilde{\psi}_j^\dagger \tilde{\psi}_j\rangle$ is the number of the non-condensate atoms on $j$-th lattice site. 
The time-dependent Hartree-Fock mean field $2U\tilde{n}_j(t)$ in Eq. (\ref{GP}) arises from the thermal gas of non-condensate atoms. 
Equation (\ref{GP}) reduces to the usual discrete Gross-Pitaevskii equation 
\cite{Inguscio1,Smerzi3,ReyBurnett} when all the atoms are assumed to be in the condensate ({\it i.e.}, if we set $\tilde{n}_j=0$).

Introducing the usual phase and amplitude variables for the condensate, $\Phi_{j}(t)=\sqrt{n_j^\mathrm{c}(t)}e^{\mathrm{i}\theta_{j}(t)}$, Eq. (\ref{GP}) is equivalent to 
\begin{eqnarray}
\frac{\mathrm d n_j^\mathrm{c}(t)}{\mathrm d t}&=&-2J\sum_{\langle l \rangle}\sqrt{n_j^\mathrm{c}(t)n_l^\mathrm{c}(t)}\sin(\theta_l(t)-\theta_j(t)),\label{continuity1}\\
\frac{\mathrm d \theta_j(t)}{\mathrm d t}
&=&J\sum_{\langle l \rangle}\sqrt{\frac{n_{l}^\mathrm{c}(t)}{n_j^\mathrm{c}(t)}}\cos(\theta_{l}(t)-\theta_{j}(t))
-U(n_j^\mathrm{c}(t)+2\tilde{n}_j(t))\nonumber\\
&\equiv&-\varepsilon_j^\mathrm{c}(t).\label{Josephson1}
\end{eqnarray}
Equation (\ref{continuity1}) is the continuity equation for the condensate and Eq. (\ref{Josephson1}) often called the Josephson equation.
The energy of a condensate atom $\varepsilon_j^\mathrm{c}(t)$ reduces 
to the equilibrium chemical potential $\mu_{\mathrm c 0}$ of the condensate in static thermal equilibrium ($\theta_j$, $n_j^\mathrm c$ and $\tilde{n}_j$ are independent of site index $j$)
\begin{eqnarray}
\mu_{\mathrm c0}=-zJ+U(n^{\mathrm c 0}(T)+2\tilde{n}^0(T)).\label{chemicalpotential}
\end{eqnarray}
Here $z$ is the number of the nearest neighbor sites, $n^{\mathrm c 0}$ is the number of the condensate atoms per site in equilibrium, and $\tilde{n}^0$ is the number of the non-condensate atoms per site in equilibrium.
The solution of Eq. (\ref{Josephson1}) in static equilibrium is 
$\theta_0(t)=-\mu_{\mathrm c0}t$.
From Eq. (\ref{continuity1}), one finds that the Josephson current between the $j$-th and $l$-th lattice sites is $J_j(t)=2J\sqrt{n_j^\mathrm{c}(t)n_l^\mathrm{c}(t)}\sin(\theta_l(t)-\theta_j(t))$.

The well-known Bogoliubov excitation spectrum for a uniform optical lattice is easily obtained from the above equations \cite{Stoof,ReyBurnett,Javanainen}. 
Considering small fluctuations from equilibrium, $n_j^\mathrm{c}(t)=n^{\mathrm{c}0}+\delta n_j^\mathrm{c}(t)$, $\theta_j(t)=\theta_0(t)+\delta\theta_j(t)$ and $\tilde{n}_j(t)=\tilde{n}^0+\delta \tilde{n}_j(t)$, Eqs. (\ref{continuity1}) and (\ref{Josephson1}) reduce to
\begin{eqnarray}
\frac{\mathrm d\delta n_j^\mathrm{c}(t)}{\mathrm d t}
&=&-2J n^{\mathrm{c}0}\sum_{\langle l\rangle}\left(\delta \theta_l(t)-\delta \theta_j(t)\right), \label{continuity2}\\
\frac{\mathrm d \delta \theta_j(t)}{\mathrm d t}&=&\frac{J}{2n^{\mathrm c 0}}\sum_{\langle l \rangle}\left(\delta n_l^\mathrm{c}(t)-\delta n_j^\mathrm{c}(t)\right)-U\left(\delta n_j^{\mathrm{c}}(t)+2\delta\tilde{n}_j(t)\right).\label{Josephson2}
\end{eqnarray}
Ignoring the non-condensate atom term ($\delta \tilde n_j=0$), the solution of these coupled equations are the normal modes
\begin{eqnarray}
\delta \theta_j(t)=\delta\theta({\bf q})e^{\mathrm{i} [{\bf q}\cdot {\bf r}_j-E_{\bf q}t]},\ \ 
\delta n_j^\mathrm{c}(t)=\delta n^\mathrm{c}({\bf q})e^{\mathrm{i} [{\bf q}\cdot{\bf r}_j-E_{\bf q}t]}.
\end{eqnarray}
The Bloch-Bogoliubov excitation energy in an optical lattice is given by
\begin{eqnarray}
E_\mathbf q&=&\sqrt{\epsilon_\mathbf q^0(\epsilon_\mathbf q^0+2Un^{\mathrm c 0})},\label{Bogoliubov}
\end{eqnarray}
where
\begin{eqnarray}
\epsilon_\mathbf q^0&\equiv &4J\sum_{i=1}^D \sin^2 \frac{q_i d}{2}.\label{freespectrum}
\end{eqnarray}
\begin{figure}
\includegraphics[width=10cm]{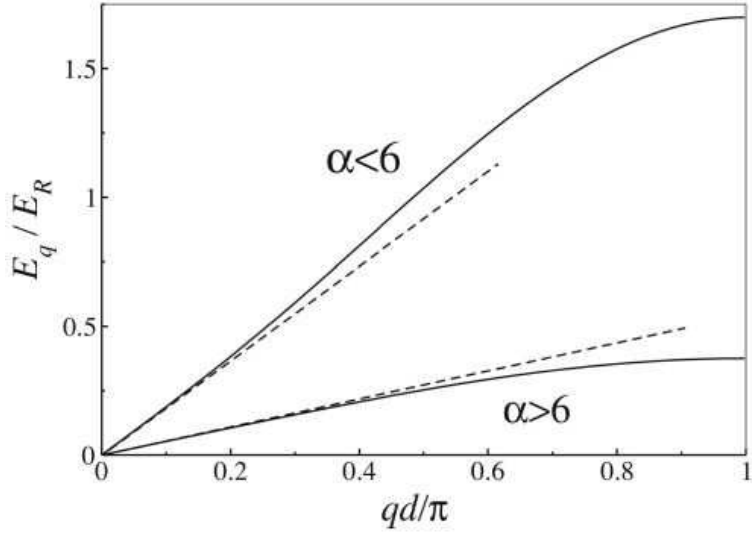}
\caption{The Bogoliubov excitation spectrum $E_q$ in a 1D optical lattice plotted as a function of the quasi-momentum $q$ in the first Brillouin zone. The spectrum is plotted for $\alpha<6$ and $\alpha>6$. The dotted lines give the low $q$ phonon dispersion relation $E_q=cq$. }
\label{dispersion}
\end{figure}
For small $q$, this spectrum is phonon-like $E_{\mathbf q}\simeq cq$, with 
the phonon velocity 
\begin{eqnarray}
c=\sqrt{2Jd^2Un^{\mathrm{c}0}}=\sqrt{\frac{Un^{\mathrm{c}0}}{m^\ast}},
\end{eqnarray}
where $m^\ast=\frac{1}{2Jd^2}$ is an effective mass of atoms in the first excitation band of the optical lattice \cite{Kramer1,Kramer2}.
This $T=0$ excitation spectrum in 1D case is shown in Fig. 1, for two values of the dimensionless interaction parameter $\alpha\equiv Un^{\mathrm c 0}/J$.

We call attention to an important feature of the dispersion
relation $E_q$ in Fig. \ref{dispersion}, considered as a function of the ratio $\alpha$. 
For $\alpha \le 6$, the excitation energy $E_q$ bends up 
before bending over, as $q$ approaches the BZ boundary.
This behavior is analogous to the so called ``anomalous dispersion'' of 
the phonon spectrum in superfluid $^4\mathrm{He}$ \cite{anomalous1,anomalous2,Pitaevskii&Levinson}.
For $\alpha>6$, in contrast, the spectrum simply bends over as one leaves the low $q$ (phonon) region. 
This feature will play a crucial role when we consider damping processes
in optical lattices.
The excitation spectrum in 2D and 3D optical lattices also exhibit this kind of spectrum.
However, the critical value of $\alpha$ then depends on the direction of ${\bf q}$, since simple cubic optical lattices in 2D and 3D do not have rotational symmetry.
The crucial effect of this anomalous dispersion on damping processes will be discussed in detail in section \ref{damping}.

\section{\label{staticPopov}static popov approximation}

In this section, we calculate the condensate fraction $n^{\mathrm c 0}/n$ and the parameter $\alpha\equiv U n^{\mathrm c 0}/J$ as a function of the optical lattice depth $s$ and the temperature. This is needed as input to the calculations in Sec. \ref{damping}.
We base our discussion on the static Popov approximation \cite{Hutchinson,ReyBurnett,WG}. 

Substituting $a_j=\Phi_j+\tilde{\psi}_j$ to $K\equiv H-\mu N$, where $N=\sum_j a_j^\dagger a_j$ is the total number of atoms, one obtains,
\begin{eqnarray}
K&=&K_0+K_1+K_2+K_3+K_4,\\
K_0&=&-J\sum_{\langle j,l\rangle}\left(\Phi_j^\ast\Phi_l+\Phi_l^\ast\Phi_j\right)
-\mu\sum_j|\Phi_j|^2+\frac{U}{2}\sum_j|\Phi_j|^4,\\
K_1&=&\sum_j\bigg(-J\sum_{\langle l\rangle}\Phi_l-\mu\Phi_j+U|\Phi_j|^2\Phi_j\bigg)\tilde{\psi}_j^\dagger\nonumber\\
& &+\bigg(-J\sum_{\langle l\rangle}\Phi_l^\ast-\mu\Phi_j^\ast+U|\Phi_j|^2\Phi_j^\ast\bigg)\tilde{\psi}_j\\
K_2&=&-J\sum_{\langle j,l\rangle}\left(\tilde{\psi}_j^\dagger\tilde{\psi}_l+\tilde{\psi}_l^\dagger\tilde{\psi}_j \right)-\mu\sum_j\tilde{\psi}_j^\dagger\tilde{\psi}_j\nonumber\\
& &+\frac{U}{2}\sum_j\left(\Phi_j^2 (\tilde{\psi}_j^\dagger)^2+4|\Phi_j|^2\tilde{\psi}_j^\dagger\tilde{\psi}_j+(\Phi_j^\ast)^2\tilde{\psi}_j^2 \right),\\
K_3&=&U\sum_j\left(\Phi_j^\ast\tilde{\psi}_j^\dagger\tilde{\psi}_j^2+\Phi_j(\tilde{\psi}_j^\dagger)^2\tilde{\psi}_j\right),\\
K_4&=&\frac{U}{2}\sum_j(\tilde{\psi}_j^\dagger)^2\tilde{\psi}_j^2.\label{K}
\end{eqnarray}
Here, we use the Hartree-Fock-Bogoliubov-Popov approximation that takes into account the third and fourth order terms of $\tilde{\psi}_j$, $\tilde{\psi}_j^\dagger$ within a mean-field approximation, but neglects terms involving the anomalous averages $\langle\tilde{\psi}_j\tilde{\psi}_j\rangle$ and $\langle\tilde{\psi}_j^\dagger\tilde{\psi}_j\tilde{\psi}_j\rangle$, namely we use
\begin{eqnarray}
K_3\simeq2U\sum_j\tilde{n}_j\left(\Phi_j^\ast \tilde{\psi}_j+\Phi_j \tilde{\psi}_j^\dagger\right),\ \ 
K_4\simeq2U\sum_j\tilde{n}_j \tilde{\psi}_j^\dagger\tilde{\psi}_j.\label{HFBP}
\end{eqnarray}

Within this approximation, we have
\begin{eqnarray}
K&=&K_0+K_1^\prime+K_2^\prime\\
K_1^\prime&=&\sum_j\bigg(-J\sum_{\langle l\rangle}\Phi_l-\mu\Phi_j+U(n_j^\mathrm c+2\tilde{n}_j)\Phi_j\bigg)\tilde{\psi}_j^\dagger\nonumber\\
& &+\bigg(-J\sum_{\langle l\rangle}\Phi_l^\ast-\mu\Phi_j^\ast+U(n_j^\mathrm c+2\tilde{n}_j)\Phi_j^\ast\bigg)\tilde{\psi}_j\label{K1}\\
K_2^\prime&=&-J\sum_{\langle j,l\rangle}\left(\tilde{\psi}_j^\dagger\tilde{\psi}_l+\tilde{\psi}_l^\dagger\tilde{\psi}_j \right)-\mu\sum_j\tilde{\psi}_j^\dagger\tilde{\psi}_j\nonumber\\
& &+\frac{U}{2}\sum_j\left(\Phi_j^2 (\tilde{\psi}_j^\dagger)^2+4n_j\tilde{\psi}_j^\dagger\tilde{\psi}_j+(\Phi_j^\ast)^2\tilde{\psi}_j^2 \right),
\end{eqnarray}
where $n_j\equiv n_j^{\mathrm c}+\tilde{n}_j=|\Phi_j|^2+\langle\tilde{\psi}_j^\dagger\tilde{\psi}_j\rangle$.
The coefficients of the linear terms in $\tilde\psi$, $\tilde\psi^\dagger$ in Eq. (\ref{K1}) identically vanish using the fact that $\Phi_j$ satisfies the generalized Gross-Pitaevskii equation Eq. (\ref{GP}). As a result, $K_1^\prime=0$. 

As discussed in Sec. \ref{intro}, we consider a Bose condensate in static thermal equilibrium in the $k=0$ Bloch state, and hence $\Phi_j=\sqrt{n^{\mathrm c 0}}$. Using the value of the condensate chemical potential in thermal equilibrium given in Eq. (\ref{chemicalpotential}), the remaining term $K_2^\prime$ in Eq. (\ref{K1}) reduces to
\begin{eqnarray}
K_2^\prime&=&-J\sum_{\langle j,l\rangle}\left(\tilde{\psi}_j^\dagger\tilde{\psi}_l+\tilde{\psi}_l^\dagger\tilde{\psi}_j \right)+(zJ+Un^{\mathrm c 0})\sum_j\tilde{\psi}_j^\dagger\tilde{\psi}_j\nonumber\\
& &+\frac{U}{2}\sum_j\left(n^{\mathrm c 0} (\tilde{\psi}_j^\dagger)^2+n^{\mathrm c 0}\tilde{\psi}_j^2 \right).\label{K2}
\end{eqnarray}
We next introduce Fourier components,
\begin{eqnarray}
\tilde\psi_j=\frac{1}{\sqrt{I^D}}\sum_{{\bf q}\neq 0}a_{\bf q} e^{\mathrm i{\bf q}\cdot {\bf r}_j}, \label{Fourier}
\end{eqnarray}
where the number of lattice sites in one direction is denoted by $I$ and hence the total number of sites is $I^D$. 
The Momentum sum is over the first Brilloiuin zone of the optical lattice.
Substituting Eq. (\ref{Fourier}) into Eq.(\ref{K2}), one finds
\begin{eqnarray}
K_2^\prime=\sum_{{\bf q}\neq0}(\epsilon_{\bf q}^0+Un^{\mathrm c 0})a_{\bf q}^\dagger a_{\bf q}+\frac{Un^{\mathrm c 0}}{2}\sum_{{\bf q}\neq0}(a_{\bf q}^\dagger a_{-\bf q}^\dagger+a_{\bf q} a_{-\bf q}).
\end{eqnarray}

We can diagonalize $K=K_0+K_2^\prime$ by the Bogoliubov transformation \cite{Stoof,ReyBurnett},
\begin{eqnarray}
\alpha_{\bf q}&=&u_{\bf q} a_{\bf q}+v_{\bf q}a_{-\bf q}^\dagger,\nonumber\\ 
\alpha_{\bf -q}^\dagger&=&v_{-\bf q}^\ast a_{\bf q}+u_{-\bf q}^\ast a_{-\bf q}^\dagger.\label{Bogoliubov_transformation}
\end{eqnarray}
If we assume that $\alpha_{\bf q}$ and $\alpha_{\bf q}^\dagger$ obey the usual Bose commutation relations, we obtain the following conditions for $u_{\bf q}$ and $v_{\bf q},$ 
\begin{eqnarray}
|u_{\bf q}|^2-|v_{\bf q}|^2=1.\label{diagonalize1}
\end{eqnarray}
From the condition for the diagonalization of $K$,
we find that $u_{\bf q}$ and $v_{\bf q}$ have to satisfy the following equation,
\begin{eqnarray}
Un^{\mathrm c 0}(u_{\bf q}^2+v_{\bf q}^2)-2(\epsilon_{\bf q}^0+Un^{\mathrm c 0})u_{\bf q}v_{\bf q}=0\label{diagonalize2}.
\end{eqnarray}
Solving both Eq. (\ref{diagonalize1}) and Eq. (\ref{diagonalize2}), one can easily derive the parameters for the Bogoliubov transformation (we take $u_{\bf q}$ and $v_{\bf q}$ real),
\begin{eqnarray}
u_{\bf q}^2=\frac{1}{2}\left(\frac{\tilde{E}_{\bf q}}{E_{\bf q}}+1\right),
v_{\bf q}^2=\frac{1}{2}\left(\frac{\tilde{E}_{\bf q}}{E_{\bf q}}-1\right).\label{Bogoliubov_parameters}
\end{eqnarray}
We have here introduced the Hartree-Fock (HF) excitation spectrum
\begin{eqnarray}
\tilde{E}_{\mathbf q}&\equiv&[\epsilon_0(q)+2U(n^{\mathrm c 0}+\tilde{n}^0)]-\mu_{\mathrm c 0}\nonumber\\ 
&=&4J\sin^2(\frac{qd}{2})+Un^{\mathrm c 0}.
\end{eqnarray}

Putting all these results together, our total Hamiltonian has been reduced to
\begin{eqnarray}
K&=&K_0+\sum_{{\bf q}\neq0}\left(\tilde{E}_{\bf q}v_{\bf q}^2-Un^{\mathrm c 0}u_{\bf q}v_{\bf q}\right)+\sum_{{\bf q}\neq0}E_{\bf q}\alpha_{\bf q}^\dagger \alpha_{\bf q}\nonumber\\
&=&-\frac{UN_{\mathrm c 0}}{2}(n^{\mathrm c 0}+4\tilde{n}^0)+\frac{1}{2}\sum_{\mathbf q \neq0}\left(E_\mathbf q-\tilde{E}_\mathbf q\right)+\sum_{\mathbf q\neq 0}E_{\mathbf q}\alpha_{\mathbf q}^\dagger \alpha_{\mathbf q}.\label{diagonalized_Hamiltonian}
\end{eqnarray}
Here $N_{\mathrm c 0}=n^{\mathrm c 0}I^D$ is the total number of condensate atoms.
The Bogoliubov-Popov excitation spectrum $E_{\mathbf q}$ appearing in Eq. (\ref{diagonalized_Hamiltonian}) is identical to Eq. (\ref{Bogoliubov}), 
except that now $n^{\mathrm c 0}(T)$ is the temperature-dependent number of 
condensate atoms in any given lattice well.
The Bogoliubov-Popov excitation spectrum can be obtained by ignoring the non-condensate fluctuation ($\delta \tilde{n}_j=0$) but using the temperature-dependent number of condensate atoms $n^{\mathrm c 0}(T)$ as given by Eq. (\ref{continuity2}) and Eq. (\ref{Josephson2}).
We thus assume that the thermal cloud is always in static equilibrium when dealing with the time-dependent density fluctuations $\delta n_j^{\mathrm c}(t)$ of the condensate. The approximation of a static thermal cloud has been successfully used in other discussions of condensate collective modes \cite{WG,Hutchinson}, which we refer the reader for further discussion.

It is easy to verify that $E_\mathbf{q}$ in Eq. (\ref{Bogoliubov}) reduces to 
the HF energy $\tilde{E}_\mathbf{q}$ in the limit $J \gg U n^{\mathrm c 0}$, {\it i.e.}, when $\alpha\ll 1$.
This HF limit corresponds to setting the Bogoliubov amplitudes $u_{\mathbf q}^2=1$, $v_{\mathbf q}^2=0$ in Eq. (\ref{Bogoliubov_parameters}).
In dealing with Bose gases trapped in harmonic potentials,
one can always use this HF approximation \cite{ZNG} for the excitations describing 
the thermal cloud as long as the kinetic energy of the atoms ($\sim k_\mathrm{B} T$) 
is much larger than the interaction energy ($U n^{\mathrm c 0}$).
In contrast, apart from the limiting case of $\alpha\ll 1$,
we must always use the {\it full} Bogoliubov spectrum $E_\mathbf{q}$ to describe the thermal cloud composed of excitations in the first band of an optical lattice.

\begin{figure}[keepaspectratio]
\includegraphics[width=9cm]{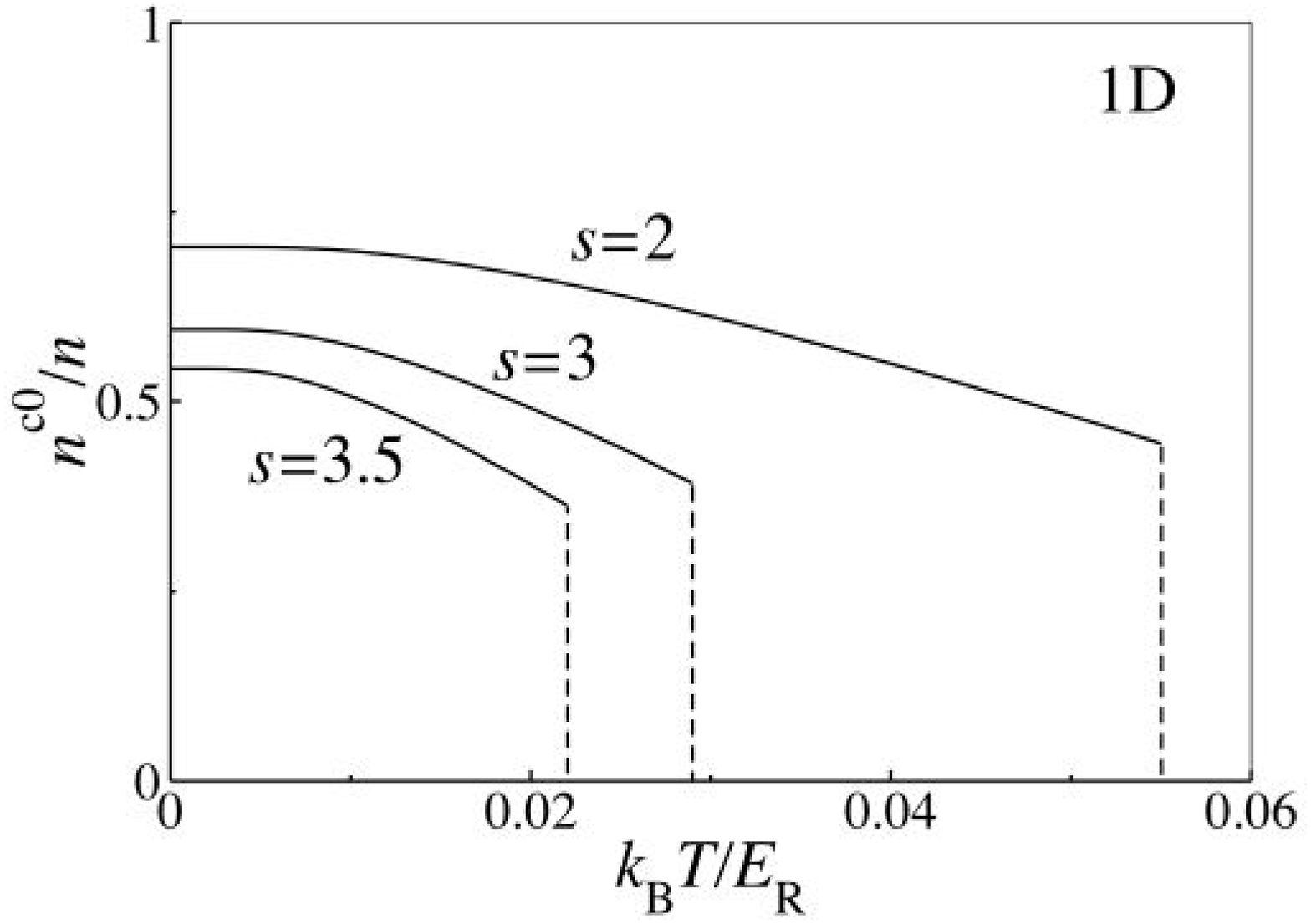}\\
\vspace{1cm}
\includegraphics[width=9cm]{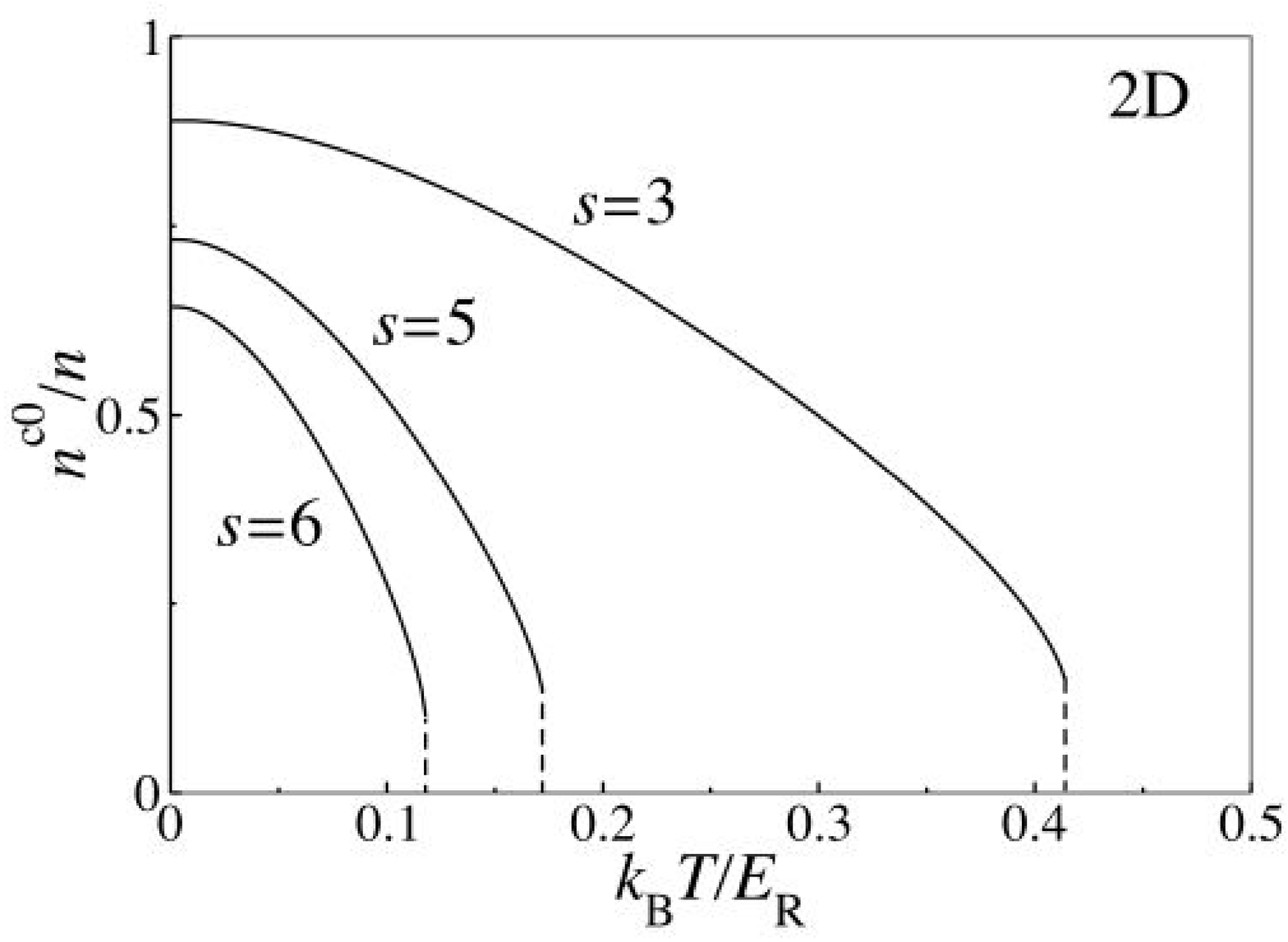}\\
\vspace{1cm}
\includegraphics[width=9cm]{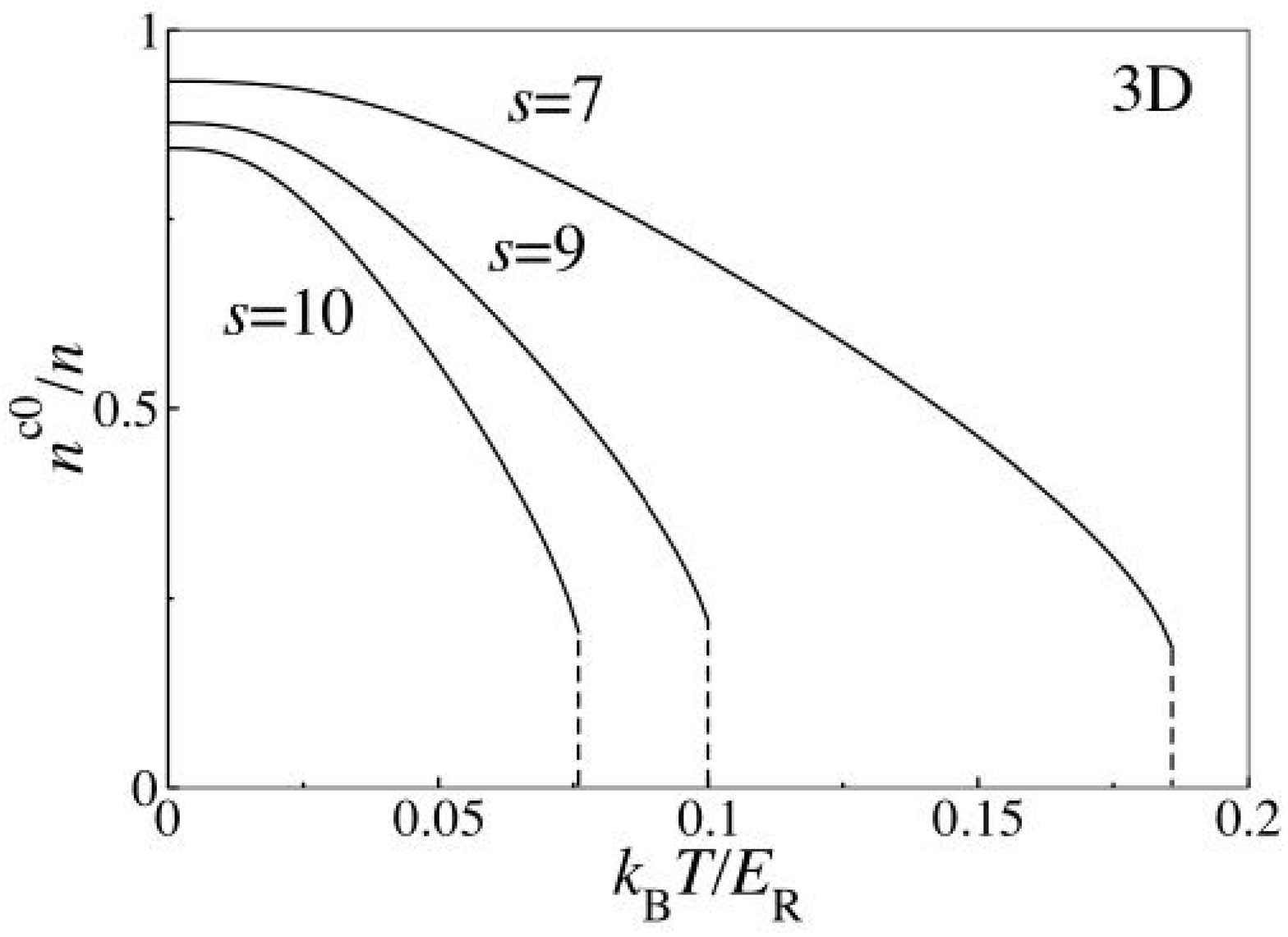}
\caption{The condensate fraction $n^{\mathrm c 0}/n$ in $D$-dimensional optical lattice as a function of temperature. In this paper, the height of the optical lattice potential (in units of $E_{\mathrm R}$) is always denoted by $s$.}
\label{condensatefraction}
\end{figure}

Expressing $\tilde{n}^0(T)$ in terms of these Bogoliubov-Popov excitations, 
we have \cite{ReyBurnett,Stoof,Hutchinson}
\begin{eqnarray}
n&=&n^{\mathrm{c}0}+\frac{1}{I^D}\sum_{{\bf q}\neq0}\langle a_{\bf q}^\dagger a_{\bf q}\rangle\nonumber\\
&=&n^{\mathrm{c}0}+\frac{1}{I^D}\sum_{{\bf q}\neq0}\left[\left(u_{\bf q}^2+v_{\bf q}^2\right)f^{0}(E_{\bf q})+v_{\bf q}^2\right],\label{selfconsist}
\end{eqnarray}
where $f^{0}(E_q)=[\exp(\beta E_q)-1]^{-1}$ is the usual Bose distribution function.
The number of condensate atoms $n^{\mathrm{c}0}$ at a site is found by solving 
Eq. (\ref{selfconsist}) self-consistently for a fixed value of the total site density $n$. The condensate fraction $n^{\mathrm c 0}/n$ in a $D$-dimensional optical lattice is shown in Fig. \ref{condensatefraction}. 
We take $n=2$ and use the parameters from Ref. \cite{Esslinger2}
The spurious finite jump in the condensate atom number $n^{\mathrm c 0}$ at the transition temperature $T_\mathrm{c}$ is an inherent problem of the Bogoliubov theory in a uniform gas (for further discussion see, for example, Ref. \cite{Shi}).

Strictly speaking, there are no solutions of Eq. (\ref{selfconsist}) at finite temperature for an infinite optical lattice in 1D and 2D because of the divergent contribution from excitations with small momentum, 
in accordance with the well-known Mermin-Wagner-Hohenberg 
theorem \cite{Mermin_Wagner}.
However, for the {\it finite} systems discussed in this paper, this divergence is not present. That is, Eq. (\ref{selfconsist}) has a solution describing a finite value of the condensate $n^{\mathrm c 0}(T)$ below $T_{\mathrm c}$. 

\begin{figure}[keepaspectratio]
\includegraphics[width=8cm]{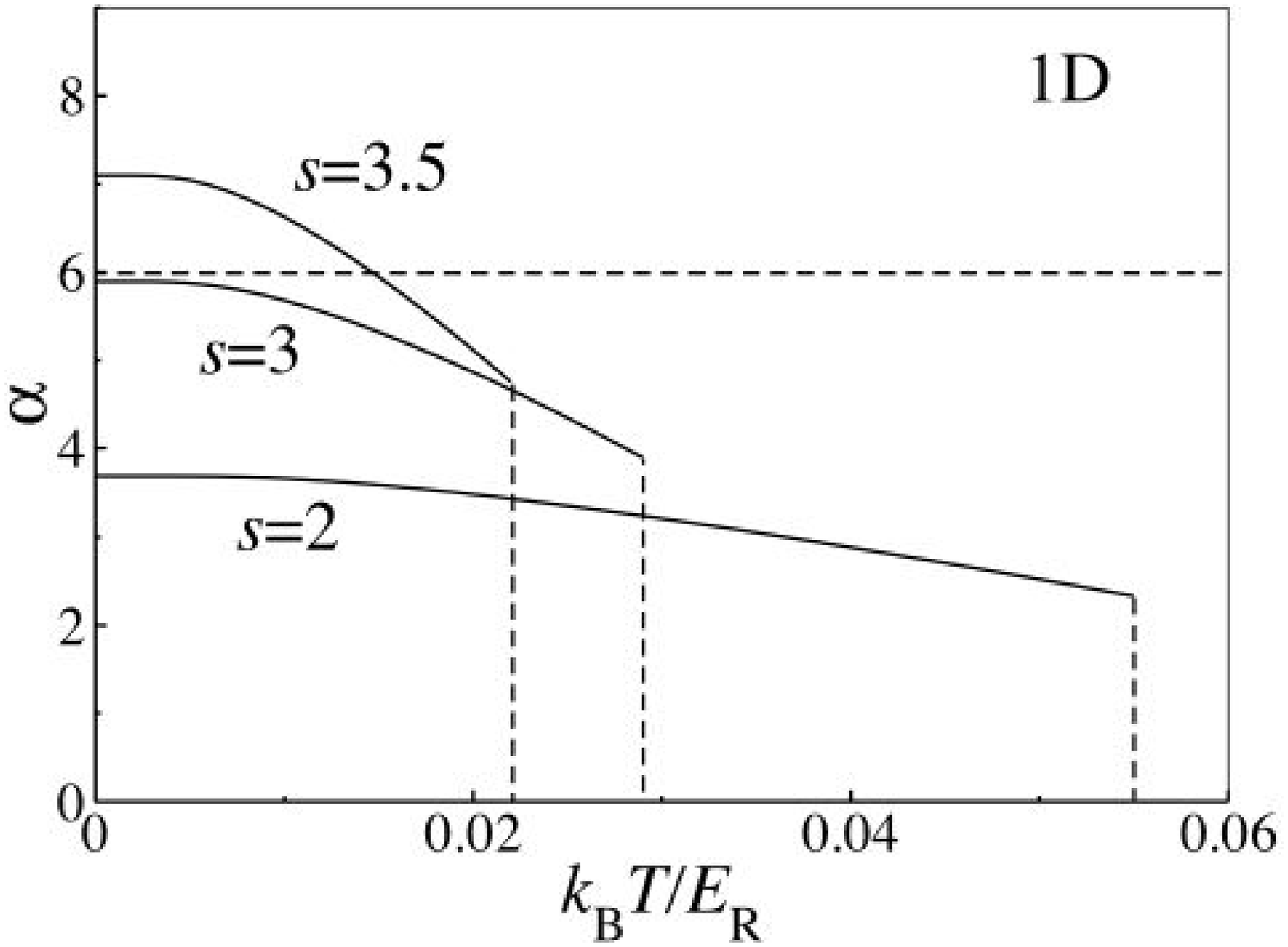}\\
\vspace{1cm}
\includegraphics[width=8cm]{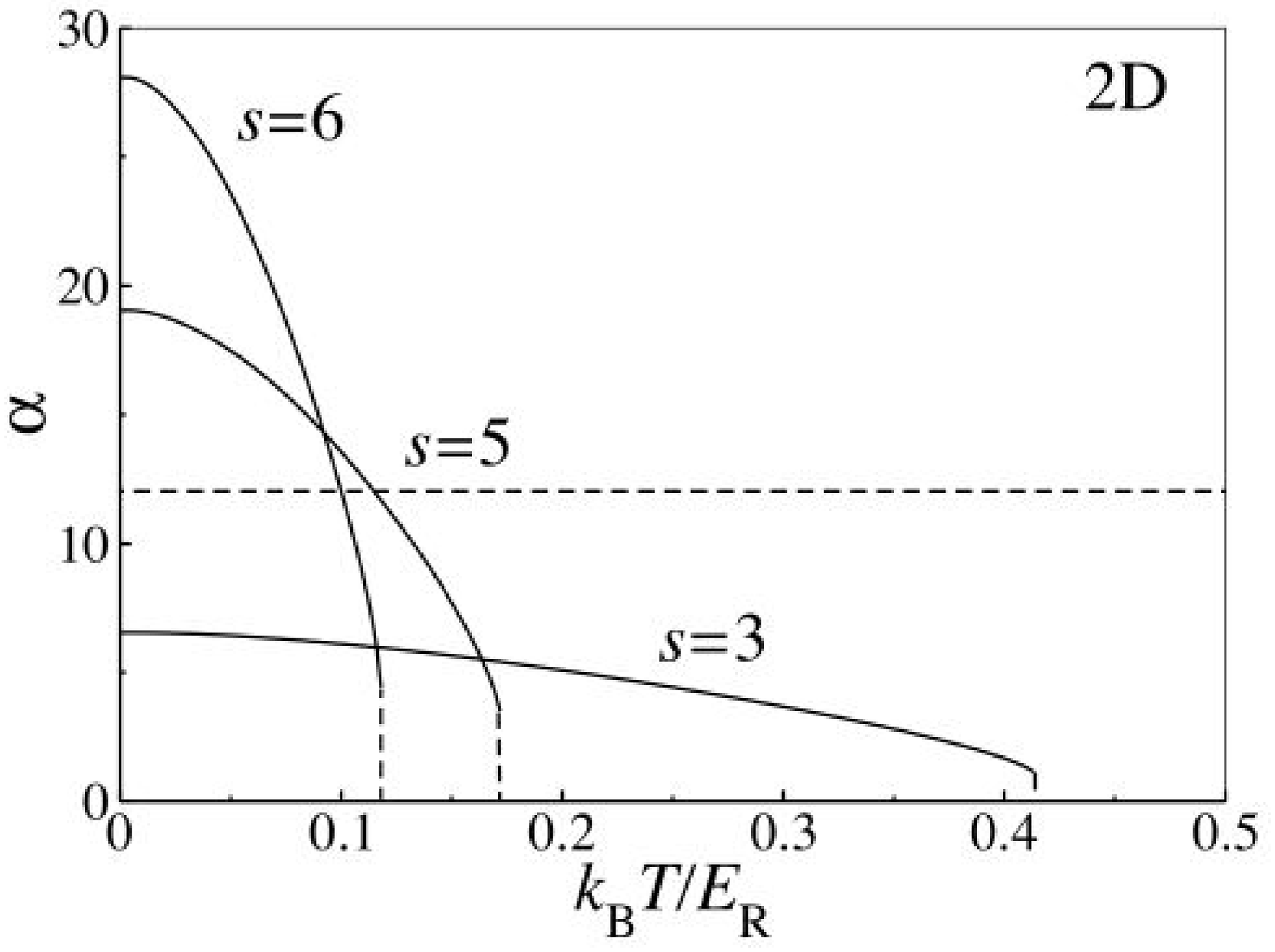}\\
\vspace{1cm}
\includegraphics[width=8cm]{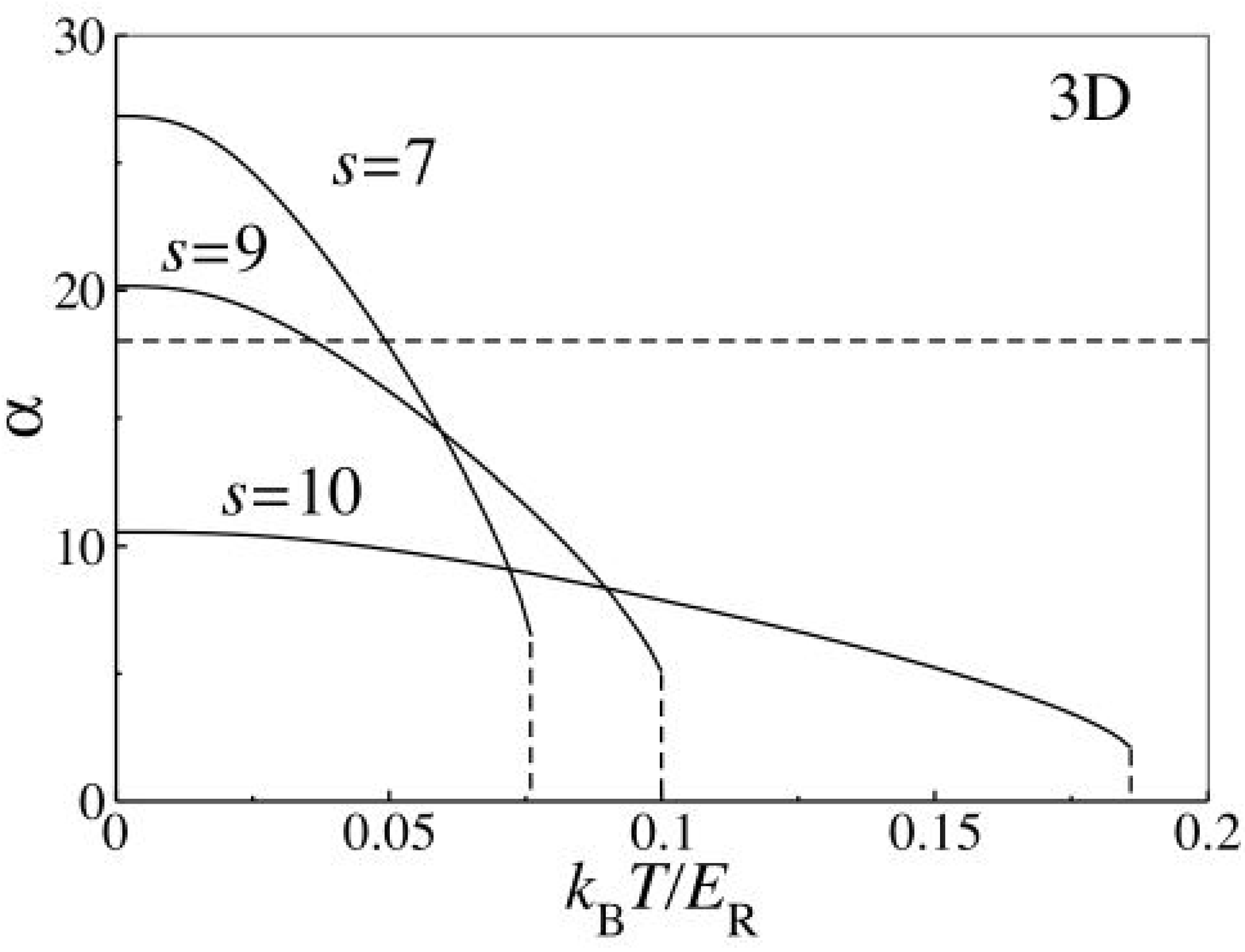}
\caption{The demensionless interaction parameter $\alpha=Un^{\mathrm c 0}(T)/J$,
plotted as a function of temperature, for several values of the optical well depth $s$.
The number of condensate atoms at a lattice site
$n^{\mathrm c 0}(T)$ is given in Fig. \ref{condensatefraction}.
The dashed line shows the critical value $\alpha_{\mathrm c}=6D$, above which there is no damping.}
\label{alpha}
\end{figure}

In Fig. \ref{alpha}, we plot the parameter $\alpha\equiv Un^{\mathrm c 0}/J$ as a function of the temperature for several values of the optical depth $s$. 
The dimensionless interaction parameter $\alpha$ and the results in Fig. \ref{alpha} will play an important role in the discussion in the rest of this paper. Since we limit our discussion to the first energy band of the optical lattice, our results only apply when $s\gg 2k_{\mathrm B} T/E_{\mathrm R}$. Higher excitation bands would be thermally populated if we consider lower values of $s$ and would have to be considered. 

\section{\label{damping}damping of Bogoliubov excitations}

In this section, we discuss the damping of condensate modes in optical lattices. 
In Sec. \ref{Green's function}, we derive a formal expression for the damping of Bogoliubov excitations using the thermal Green's function formalism. 
In Sec. \ref{1D}, we use this expression to calculate the Landau damping of Bogoliubov excitations in 1D lattice. Section \ref{2and3D} extends this analysis to 2D and 3D optical lattices.

\subsection{\label{Green's function} Green's function technique}
In this section, we take $K_{\mathrm B}\equiv K_0+K_1+K_2$ in Eq. (\ref{K}) as the bare zeroth-order Hamiltonian and treat $K_3$ as a perturbation 
(We neglect the fourth order term $K_4$). 
We use the simplest Bogoliubov theory to diagonalize $K_{\mathrm B}$ \cite{Stoof,ReyBurnett}. 
We neglect the fourth order term in the fluctuation $K_4$.
The chemical potential for $K_B$ is given by $\mu_{\mathrm B 0}=-zJ+Un^{\mathrm c 0}$ (from the condition $K_1=0$).
$K_2$ can be diagonalized by the Bogoliubov transformation Eq. (\ref{Bogoliubov_transformation}) and one obtains Eq. (\ref{diagonalized_Hamiltonian}).

Using Eq. (\ref{Fourier}) and Eq. (\ref{Bogoliubov_transformation}), $K_3$ can be re-written in terms of the Bogoliubov quasiparticles 
\begin{eqnarray}
K_3&=&\frac{1}{2}\sum_{{\bf p}_1,{\bf p}_2,{\bf p}_3\neq0}M_{{\bf p}_1,{\bf p}_2;{\bf p}_3}\left(\alpha_{{\bf p}_3}^\dagger\alpha_{{\bf p}_1}\alpha_{{\bf p}_2}+ \alpha_{{\bf p}_3}\alpha_{{\bf p}_1}^\dagger\alpha_{{\bf p}_2}^\dagger\right)\nonumber\\
& &+\sum_{{\bf p}_1,{\bf p}_2,{\bf p}_3\neq0}L_{{\bf p}_1,{\bf p}_2;{\bf p}_3}\left(\alpha_{{\bf p}_3}^\dagger\alpha_{-{\bf p}_1}^\dagger\alpha_{-{\bf p}_2}^\dagger+ \alpha_{-{\bf p}_3}\alpha_{{\bf p}_1}\alpha_{{\bf p}_2}\right),
\end{eqnarray}
where the matrix elements are
\begin{eqnarray}
M_{{\bf p}_1,{\bf p}_2;{\bf p}_3}&=&2U\sqrt{\frac{n^{\mathrm c 0}}{I^D}}\sum_{\bf G}\big[(u_{{\bf p}_1}u_{{\bf p}_3}+v_{{\bf p}_1}v_{{\bf p}_3}-v_{{\bf p}_1}u_{{\bf p}_3})u_{{\bf p}_2}\nonumber\\
& &-(u_{{\bf p}_1}u_{{\bf p}_3}+v_{{\bf p}_1}v_{{\bf p}_3}-u_{{\bf p}_1}v_{{\bf p}_3})v_{{\bf p}_2}\big]
\delta_{{\bf p}_1+{\bf p}_2, {\bf p}_3+{\bf G}},\label{matrixelement1}\\
L_{{\bf p}_1,{\bf p}_2;{\bf p}_3}&=&U\sqrt{\frac{n^{\mathrm c 0}}{I^D}}\sum_{\bf G}\left(v_{{\bf p}_1} v_{{\bf p}_2}u_{{\bf p}_3}-u_{{\bf p}_1}u_{{\bf p}_2}v_{{\bf p}_3}\right)\delta_{{\bf p}_1+{\bf p}_2, {\bf p}_3+{\bf G}},\label{matrixelement2}
\end{eqnarray}
and ${\bf G}$ is a reciprocal lattice vector. 
The Kronecker delta $\delta_{{\bf p}_1+{\bf p}_2, {\bf p}_3+{\bf G}}$ 
expresses the conservation of quasi-momentum of the three excitation 
scattering processes (Umklapp processes are associated with $\mathbf G\neq 0$).

The thermal Green's function for Bogoliubov quasiparticles is defined by (see e.g. \cite{Mahan})
\begin{eqnarray}
G_{\bf q}(\tau_1-\tau_2)\equiv-\langle T_\tau\left(\alpha_{\bf q}(\tau_1)\alpha_{\bf q}^\dagger(\tau_2)\right)\rangle,
\end{eqnarray}
where 
\begin{eqnarray}
\alpha_{\bf q}(\tau)\equiv e^{K\tau}\alpha_{\bf q}e^{-K\tau},\ \ 
\alpha_{\bf q}^\dagger(\tau)\equiv e^{K\tau}\alpha_{\bf q}^\dagger e^{-K\tau}.
\end{eqnarray}
The angular brackets $\langle...\rangle $ indicates taking an average for thermal equilibrium and $T_\tau$ gives the usual time ordered product of operators.
Fourier transformation of this imaginary time Green's function is given by
\begin{eqnarray}
G_{\bf q}(\tau)=\frac{1}{\beta}\sum_{\omega_n}G_{\bf q}(\mathrm i \omega_n)e^{-\mathrm i \omega_n\tau},
\end{eqnarray}
where $\omega_n$ is the boson Matsubara frequency $\omega_n\equiv\frac{2n\pi}{\beta}$ ($n$ is an integer).
The zero-th order Green's function is 
\begin{eqnarray}
G_{\bf q}^0(\mathrm i \omega_n)=\frac{1}{\mathrm i \omega_n-E_{\bf q}},
\end{eqnarray}
where $E_{\bf q}$ is the Bloch-Bogoliubov excitation energy given in Eq. (\ref{Bogoliubov}).

The second-order self-energy terms are given by the usual diagrams shown in Fig. 4. 
\begin{figure}[keepaspectratio]
\includegraphics[width=10cm]{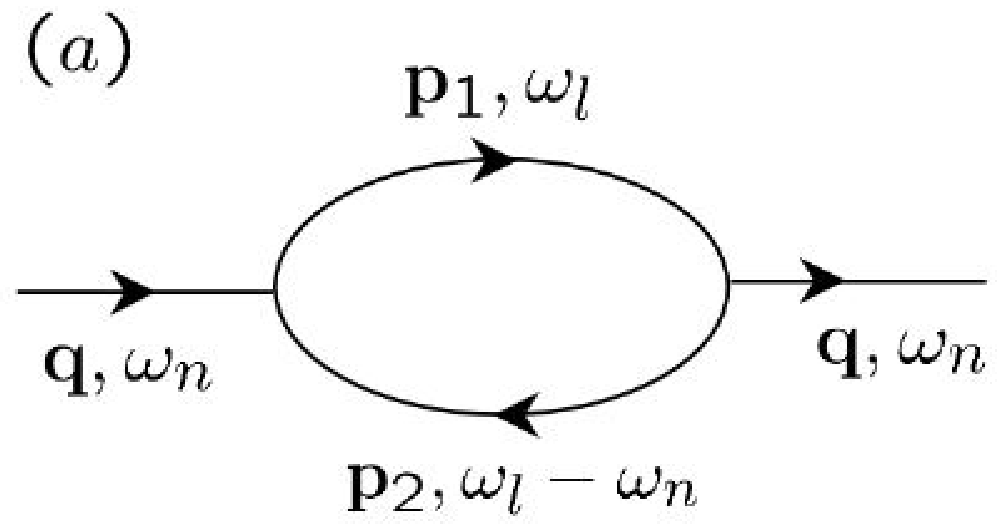}
\includegraphics[width=10cm]{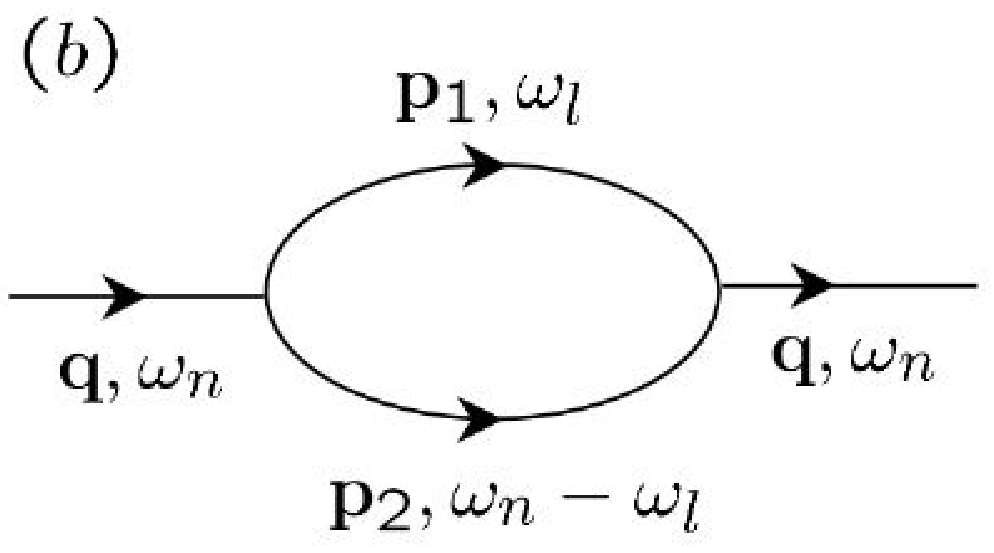}
\caption{The second order self-energys.}
\label{diagram}
\end{figure}
Fig. \ref{diagram} (a) describes a condensate excitation of (quasi) momentum ${\bf q}$ being absorbed by an excitation ${\bf p}_2$ of the optical lattice thermal gas, lending to a thermal excitation with momentum ${\bf p}_1$. 
This describes Landau damping processes \cite{Szepfalusy, PitaevskiiStringari,Fedichev}.
Fig. \ref{diagram} (b) describes a condensate excitation of momentum ${\bf q}$ decay into two excitations with momentums ${\bf p}_1$ and ${\bf p}_2$.
This describes Beliaev damping \cite{Beliaev}.
These diagrams give the second-order self-energy of the Bogoliubov excitation  
\begin{eqnarray}
\Sigma_{\bf q}(\mathrm i\omega_n)&=&\Sigma^{\mathrm L}_{\bf q}(\mathrm i\omega_n)+\Sigma^{\mathrm B}_{\bf q}(\mathrm i\omega_n)\nonumber,\\
\Sigma^{\mathrm L}_{\bf q}(\mathrm i\omega_n)&=&-\frac{1}{\beta}\sum_{{\bf p}_1,{\bf p}_2\neq 0}\sum_{\omega_l}|M_{{\bf q},{\bf p}_2;{\bf p}_1}|^2 G_{{\bf p}_1}^0(\mathrm i \omega_l)G_{{\bf p}_2}^0(\mathrm i\omega_l-\mathrm i\omega_n),\nonumber\\
\Sigma^{\mathrm B}_{\bf q}(\mathrm i\omega_n)&=&-\frac{1}{2\beta}\sum_{{\bf p}_1,{\bf p}_2\neq 0}\sum_{\omega_l}|M_{{\bf p}_1,{\bf p}_2;{\bf q}}|^2 G_{{\bf p}_1}^0(\mathrm i\omega_l)G_{{\bf p}_2}^0(\mathrm i\omega_n-\mathrm i\omega_l).
\end{eqnarray}
The summation over Matsubara frequencies is calculated by the well-known formula (see Ref. \cite{Mahan})
\begin{eqnarray}
\sum_{\omega_n}g(\omega_n)=-\frac{\beta}{2\pi\mathrm i}\oint_C\mathrm dz g(z)f^0(z),\label{Matsubarasum}
\end{eqnarray}
where $f^0(z)$ is the Bose distribution function and the contour encircles the poles of $g(z)$.
Using Eq. (\ref{Matsubarasum}), one obtains
\begin{eqnarray}
\Sigma^{\mathrm L}_{\bf q}(\mathrm i\omega_n)&=&-\sum_{{\bf p}_1,{\bf p}_2\neq 0}|M_{{\bf q},{\bf p}_2;{\bf p}_1}|^2\frac{f^0(E_{{\bf p}_1})-f^0(E_{{\bf p}_2})}{\mathrm i\omega_n-(E_{{\bf p}_1}-E_{{\bf p}_2})},\\
\Sigma^{\mathrm B}_{\bf q}(\mathrm i\omega_n)&=&-\frac{1}{2}\sum_{{\bf p}_1,{\bf p}_2\neq 0}|M_{{\bf p}_1,{\bf p}_2;{\bf q}}|^2\frac{1+f^0(E_{{\bf p}_1})+f^0(E_{{\bf p}_2})}{\mathrm i\omega_n-E_{{\bf p}_1}-E_{{\bf p}_2}}.
\end{eqnarray}

The damping of excitations is given by the imaginary part of the self-energy
\begin{eqnarray}
\Gamma_{\bf q}=-\mathrm{Im}\Sigma_{\bf q}(E_{\bf q}+\mathrm i\delta),
\end{eqnarray}
where $\delta\to +0$. This gives the damping of Bogoliubov excitations in an optical lattice,
\begin{eqnarray}
\Gamma_{\bf q}&=&\Gamma^{\mathrm L}_{\bf q}+\Gamma^{\mathrm B}_{\bf q},\\
\Gamma^{\mathrm L}_{\bf q}&=&\pi \sum_{{\bf p}_1,{\bf p}_2\neq 0}|M_{{\bf q},{\bf p}_2;{\bf p}_1}|^2[f^0\left(E_{{\bf p}_2}\right)-f^0\left(E_{{\bf p}_1}\right)]\delta\left(E_{ \bf q}-E_{{\bf p}_1}+E_{{\bf p}_2}\right),\label{Landaudamp}\\
\Gamma^{\mathrm B}_{\bf q}&=&\frac{\pi}{2}\sum_{{\bf p}_1,{\bf p}_2\neq 0}|M_{{\bf p}_1,{\bf p}_2;{\bf q}}|^2[1+f^0(E_{{\bf p}_1})+f^0(E_{{\bf p}_2})]\delta\left(E_{ \bf q}-E_{{\bf p}_1}-E_{{\bf p}_2} \right).\label{Beliaevdamp}
\end{eqnarray}
Landau damping $\Gamma^{\mathrm L}_{\bf q}$ is expected to be dominant at higher temperatures where there are thermally excited quasiparticles.
In contrast, Beliaev damping $\Gamma^{\mathrm B}_{\bf q}$ is due to a decay process and can arise in the absence of thermally-excited excitations.
Beliaev damping is possible even at $T=0$ [{\it i.e.} $f^0(E_{\bf p})=0$] and is expected to be dominant at low temperatures.

\subsection{\label{1D}Landau damping in a 1D optical lattice}

\begin{figure}[keepaspectratio]
\includegraphics[width=10cm]{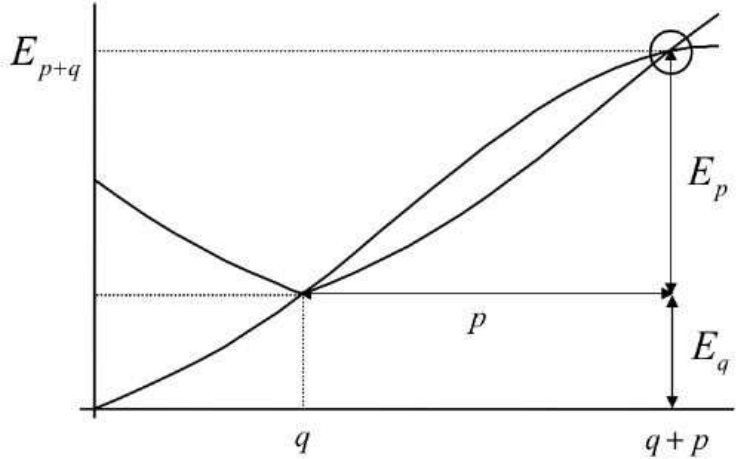}
\caption{The Bogoliubov excitation energy $E_q$ in a 1D optical lattice, for $\alpha<6$. The intersection of the two dispersion curves at $p+q$ shows that the energy conservation condition is satisfied in this case.}
\vspace{2cm}
\label{Peierls}
\end{figure}

In this subsection, we calculate the Landau damping given by Eq. (\ref{Landaudamp}) for a 1D optical lattice.
For this, the energy conservation condition $E_{\bf q}+E_{{\bf p}}=E_{{\bf q}+{\bf p}+{\bf G}}=E_{{\bf q}+{\bf p}}$ needs to be satisfied, where $\bf G$ is a reciprocal lattice vector.
The solution of the energy conservation condition $E_q+E_p=E_{q+p}$ for a 1D optical lattice is illustrated in Fig. \ref{Peierls} \cite{Peierls}.
First, we draw an excitation spectrum and mark $q$ on the line. 
Then we draw the spectrum line again with $q$ as origin.
If those two lines intersect, solutions consistent with energy conservation condition for three excitation process exist. The intersection can be taken as $(q+p,E_q+E_p=E_{q+p})$.
Clearly, the condition for these two dispersion curves to have intersections requires that the dispersion relation $E_q$ first bends up as $q$ increases, before bending over.
From Fig. 1, we see that the 1D optical lattice dispersion relation $E_q$ has this feature only for $\alpha< 6$ and thus Landau damping can occur only when $\alpha< 6$.
If the intersection $(q+p,E_q+E_p)$ is outside of the first Brillouin zone, 
it has to be reduced in the first Brillouin zone by subtracting a reciprocal lattice vector $G_n=\frac{2\pi n}{d}$ ($n$ is an integer), corresponding to an Umklapp process \cite{Peierls}.

\begin{figure}[keepaspectratio]
\includegraphics[width=10cm]{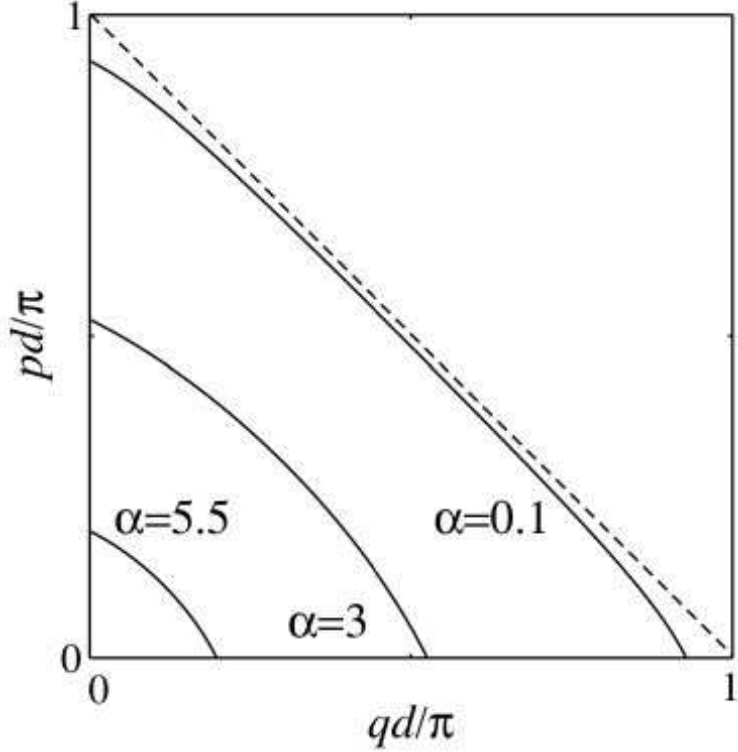}
\caption{The values $(q,p)$ satisfying the energy conservation condition $E_q+E_p=E_{q+p}$, for several values of the interaction parameter $\alpha$.}
\label{energycon1D_solution}
\end{figure}

The values of $(q,p)$ satisfying the energy conservation condition for 1D optical lattice are shown in Fig. \ref{energycon1D_solution}.
For a given value of $q$, we see that as $\alpha\to 6$, the value of $p$ decreases to zero. 
There is no solution for $\alpha>6$, indicating the disappearance of the damping for an excitation $E_q$  with any value of $q$. 
In Fig. 5, the curves of the solution of the energy conservation condition 
never cross the dashed line at $q+p=\pi/d$. 
Thus Umklapp scattering processes ($G_n\neq 0$) do not contribute to Landau damping in our present single-band model.
 
When the excitation of wavevector $q$ has a wavelength much larger than the thermal excitation $p$ ({\it i.e.}, when $q\ll p$, $\pi/d$),
one finds that $\frac{\mathrm d E_{p}}{\mathrm d p}=c$ from the energy conservation condition $E_q+E_{p}=E_{q+p}$ in Eq. (\ref{Landaudamp}). 
The Landau damping of the excitation $E_q=cq$ comes from absorbing a thermal excitation $E_{p}$ with a group velocity equal to $c$ \cite{Pitaevskii&Levinson}. This condition requires
\begin{eqnarray}
\frac{\mathrm d E_{p}}{\mathrm d p}&=&\frac{\tilde{E_p}}{E_p}4Jd\sin{\frac{pd}{2}}\cos\frac{pd}{2}\nonumber\\
&=&c\equiv\sqrt{2Jd^2Un^{\mathrm c 0}}.\label{energycon}
\end{eqnarray}
This equation can be written in the form
\begin{eqnarray}
16X^2+8(\alpha-2)X+(\alpha^2-6\alpha)=0,
\end{eqnarray}
where $X\equiv \sin^2\frac{pd}{2}$.
The unique value of $p_0$ such that $\frac{\mathrm d E_{p_0}}{\mathrm d p_0}$ in Eq. (\ref{energycon}) equals $c$  
is found to be given by the condition 
\begin{eqnarray}
\sin^2\left(\frac{p_0d}{2}\right)=\frac{-(\alpha-2)+\sqrt{2(\alpha+2)}}{4}.
\label{p_0}
\end{eqnarray}
This expression is only valid for $\alpha$ smaller than 6, such that $p_0\gg q$.

For $q\ll p$, where we can use the following approximations,
\begin{eqnarray}
u_q&\simeq&\sqrt{\frac{Un^{\mathrm c 0}}{2cp}}\left(1+\frac{cp}{2Un^{\mathrm c 0}} \right),\\
v_q&\simeq&\sqrt{\frac{Un^{\mathrm c 0}}{2cp}}\left(1-\frac{cp}{2Un^{\mathrm c 0}} \right),\\
u_{q+p}&\simeq& u_p-\frac{J\sin(pd)}{2u_p}\frac{(Un^{\mathrm c 0})^2}{E_p^3}qd,\\
v_{q+p}&\simeq& v_p-\frac{J\sin(pd)}{2v_p}\frac{(Un^{\mathrm c 0})^2}{E_p^3}qd,
\end{eqnarray}
the matrix element in Eq. (\ref{Landaudamp}) reduces to
\begin{eqnarray}
M_{q,p;q+p}&=&2U\sqrt{\frac{n^{\mathrm c 0}}{I}}\big[\left(u_p u_{q+p}+v_p v_{q+p}-v_p u_{q+p}\right)u_q-\left(u_p u_{q+p}+v_p v_{q+p}-u_p v_{q+p}\right)v_q\big]\nonumber\\
&\simeq&2U\sqrt{\frac{n^{\mathrm c 0}}{I}}\left(\frac{\epsilon^0_p}{E_p}+\frac{Un^{\mathrm c 0}}{2E_p}-\frac{(Un^{\mathrm c 0})^2}{E_p^2}\frac{Jd}{c}\sin pd \right)\sqrt{\frac{cq}{2Un^{\mathrm c 0}}}.
\end{eqnarray}
From the energy conservation condition
\begin{eqnarray}
c=\frac{\mathrm d E_p}{\mathrm d p}=\frac{\tilde{E}_p}{E_p}\ 2Jd\sin pd,
\end{eqnarray}
the matrix element can be approximated by
\begin{eqnarray}
M_{q,p;q+p}=U\left(\frac{\epsilon_p^0}{E_p}+\frac{E_p}{\tilde{E}_p}\right)\left(\frac{q}{2m^\ast c}\right)^{1/2}\frac{N_{\mathrm c 0}^{1/2}}{I},\label{matrixelement1*}
\end{eqnarray}
where $N_{\mathrm c 0}\equiv n^{\mathrm c 0}I$. This is the analogue of Eq. (10.76) in Ref. \cite{Pethick&Smith}.
Using the delta function for energy conservation 
to integrate over $p$, the Landau damping is finally reduced to
\begin{eqnarray}
\Gamma^\mathrm L_q=\frac{\alpha}{8(\alpha+2)}\left(\frac{E_{p_0}}{\epsilon^0_{p_0}} \right)^3\left(\frac{\epsilon^0_{p_0}}{E_{p_0}}+\frac{E_{p_0}}{\tilde{E}_{p_0}} \right)^2\frac{\beta J Uqd}{\sinh^2\frac{\beta E_{p_0}}{2}}.\hspace{0.3cm}\label{Landaudamping}
\end{eqnarray}

One finds there is no damping when $\alpha>6$ (see Fig. 3 for $\alpha$ as a function of $s$ and $T$). In fact, $\Gamma^\mathrm L_q$ diverges at $\alpha=6$, but this is due to the 1D nature of our system. This divergence does not occur in 2D and 3D lattices.


\subsection{\label{2and3D}Landau damping in 2D and 3D optical lattices}

We next discuss the energy conservation condition in a 2D optical lattice. 
We imagine a surface in a three dimensional space which satisfies the Bogoliubov dispersion relation $(q_x,q_y,E_{\mathbf q})$.
Then, we draw a new dispersion $E_{\mathbf q}$ in the three dimensional space, with a point on the surface $(q_{1x},q_{1y},E_{{\mathbf q}_1})$ as origin.
That is, we draw $(q_x,q_y,E_{{\mathbf q}-{\mathbf q}_1}+E_{{\mathbf q}_1})$.
If those two surfaces intersect, the energy conservation condition $E_{{\mathbf q}_1}+E_{{\mathbf q}_2}=E_{{\mathbf q}_1+{\mathbf q}_2}$ is satisfied, the intersection being given by $(q_{1x}+q_{2x},q_{1x}+q_{2y},E_{{\mathbf q}_1+{\mathbf q}_2})$. 
For this condition to be satisfied, the surface $(q_x,q_y,E_{\mathbf q})$ has to be above the other surface $(q_x,q_y,E_{\mathbf q-{\mathbf q}_1}+E_{\mathbf q})$ around $(q_{1x},q_{1y},E_{{\mathbf q}_1})$.
Since the Bogoliubov spectrum is phonon like $E_{\bf q}\simeq c q$ for small $q$, the maximum gradient of $E_{\bf q}$ at $(q_{1x},q_{1y},E_{{\bf q}_1})$ must be greater than $c$.
Since $|\nabla_{\bf q}E_{\bf q}|_{{\bf q}={\bf q}_1}$ is the maximum gradient of $E_{\bf q}$ at ${\bf q}_1$, this condition is equivalent to the requirement:
\begin{eqnarray}
|\nabla_{\bf q}E_{\bf q}|_{{\bf q}={\bf q}_1}=2Jd\frac{\tilde{E_{{\bf q}_1}}}{E_{{\bf q}_1}}\sqrt{\sin^2q_{1x}d+\sin^2q_{1y}d}>c.\label{anomalous2D}
\end{eqnarray}
Equation (\ref{anomalous2D}) is the 2D version of the condition for the Bogoliubov spectrum of a 1D optical lattice to have anomalous dispersion.
If Eq. (\ref{anomalous2D}) is satisfied, the energy conservation condition $E_{\mathbf q_1}+E_{\mathbf q_2}=E_{{\mathbf q}_1+{\mathbf q}_2}$ can be satisfied. An excitation $E_{{\mathbf q}_1}$ can then decay into $E_{{\mathbf q}_1+{\mathbf q}_2}$ by absorbing $E_{{\mathbf q}_2}$ (Landau), or an excitation $E_{{\mathbf q}_1+{\mathbf q}_2}$ can decay into two excitations $E_{{\mathbf q}_1}$ and $E_{{\mathbf q}_2}$ (Beliaev). 
 
The condition for such a ${\bf q}_1$ to exist is that the maximum value of $|\nabla_{\bf q}E_{\bf q}|$ as a function of $(q_x,q_y)$ is greater than $c$. 
Due to $|\nabla_{\bf q}E_{\bf q}|_{{\bf q}=0}=c$, we only need to consider the condition that $|\nabla_{\bf q}E_{\bf q}|$ takes its maximum at ${\bf q}\neq0$.
When $|\nabla_{\bf q}E_{\bf q}|$ has its maximum value,    
\begin{eqnarray}
\nabla_{\bf q}|\nabla_{\bf q}E_{\bf q}|&=&\frac{2Jd^2\sin q_xd}{\sqrt{\sin^2 q_xd+\sin^2 q_yd}}\nonumber\\
& &\times \left(\frac{\tilde{E}_{\bf q}}{E_{\bf q}}
\left(
\begin{array}{c}
\cos q_xd\\
\cos q_yd
\end{array}
\right)
-\frac{2JU^2(n^{\mathrm c0})^2}{E_{\bf q}^3}(\sin^2q_xd+\sin^2q_yd)
\left(
\begin{array}{c}
1\\
1
\end{array}
 \right)
\right)=0.\label{nabla2E}
\end{eqnarray}
From Eq. (\ref{nabla2E}), $|\nabla_{\bf q}E_{\bf q}|$ has its maximum value when $\cos q_xd=\cos q_y d$, {\it i.e.}, at the two values $q_x=\pm q_y$.
If we assume $q_x=q_y$ and define $u\equiv \sin^2q_xd$, Eq. (\ref{nabla2E}) reduces to
\begin{eqnarray}
u=\sin^2q_xd=\frac{-(3\alpha-4)+\sqrt{5\alpha^2+24\alpha+16}}{16}.\label{maxu}
\end{eqnarray}
From Eq. (\ref{maxu}), one sees that $u$ decreases as $\alpha$ increases, vanishing when $\alpha=12$.
Therefore, we conclude that the Landau damping when $\alpha\alt 12$ is due to excitations with momentum $q_x=\pm q_y$, and all damping processes in a 2D optical lattice will
vanish when $\alpha>12$. We have confirmed this analytical result by numerically solving the energy conservation condition.

The condition for the disappearance of Landau damping in a 3D optical lattice can be derived by generalizing the procedure described above for a 2D optical lattice. 
One sees that $|\nabla_{\bf q}E_{\bf q}|$ has its maximum when $\cos q_x d=\cos q_y d=\cos q_z d$, {\it i.e.}, $q_x=\pm q_y=\pm q_z$ and
\begin{eqnarray}
\sin^2q_xd=\frac{-3(\alpha-2)+\sqrt{5\alpha^2+36\alpha+36}}{24}.
\end{eqnarray}
One finds that $u\to 0$ when $\alpha\to 18$, and damping in a 3D optical lattice vanishes when $\alpha\ge 18$. As in the 2D case, Landau damping when $\alpha\alt 18$ only occurs for an excitation with momentum $q_x=\pm q_y=\pm q_z$.

We discuss the energy conservation condition in a 2D optical lattice in detail.
We only consider the damping of long wavelength phonon $\mathbf q$.
Using the approximation for the long wavelength phonon $E_\mathbf q\sim cq$ and
$E_{\mathbf{q}+\mathbf{p}}\sim E_{\mathbf q}+\nabla_{\mathbf p}E_{\mathbf p}\cdot \mathbf q$,
the energy conservation condition $E_{\mathbf{q}}+E_{\mathbf{q}}=E_{\mathbf{q}+\mathbf{p}}$ can be written as
\begin{eqnarray}
\frac{\alpha}{2}\left(q_x^2+q_y^2\right)=\frac{\tilde{E}_{\mathbf p}^2}{E_{\mathbf p}^2}\left(\sin^2(p_xd) q_x^2 +2\sin(p_xd)\sin(p_yd)q_xq_y+ \sin^2(p_yd) q_y^2 \right).\label{energyconserv2D}
\end{eqnarray}
When $q_x>0$ and $q_y=0$, Eq. (\ref{energyconserv2D}) can be solved easily. Defining $X\equiv\sin^2 \frac{p_xd}{2}$ and $Y\equiv\sin^2\frac{p_yd}{2}$,
the solution of Eq. (\ref{energyconserv2D}) is
\begin{eqnarray}
Y=-\left(X+\frac{\alpha}{4} \right)+\frac{1}{4}\sqrt{\frac{\alpha^3}{8X^2-8X+\alpha}}.\label{energyconserv2D_solution}
\end{eqnarray}
One can confirm that Eq. (\ref{energyconserv2D_solution}) reduces to the 1D result in Eq. (\ref{p_0}) when $Y=0$.

\begin{figure}
\includegraphics[width=10cm]{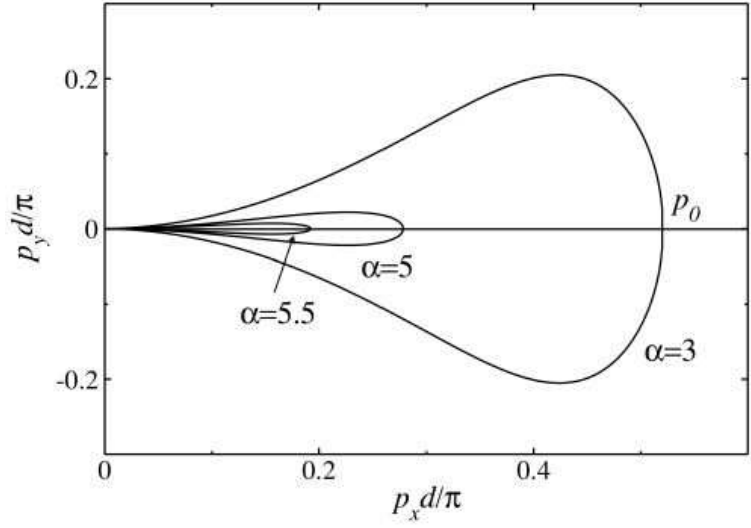}
\caption{The solution of the energy conservation condition $E_{\mathbf q}+E_{\mathbf p}=E_{\mathbf q+\mathbf p}$ for a 2D optical lattice, with $q_x>0$ and $q_y=0$. $p_0$ is defined in Eq. (\ref{p_0}) and we note $p_0\to 0$ as $\alpha\to 6$. }
\label{energyconserv2D_qy=0}
\end{figure}

Equation (\ref{energyconserv2D_solution}) is plotted in Fig. \ref{energyconserv2D_qy=0} for several values of $\alpha$. We see that as $\alpha\to 6$ the line in the $(p_x,p_y)$ plane which satisfies the energy conservation condition shrinks and vanishes when $\alpha>6$. Therefore, the Landau damping of an excitation with $q_y=0$ disappears when $\alpha>6$.
$p_0$ in Fig. 2.8 is given by Eq. (\ref{p_0}).

\begin{figure}
\includegraphics[width=10cm]{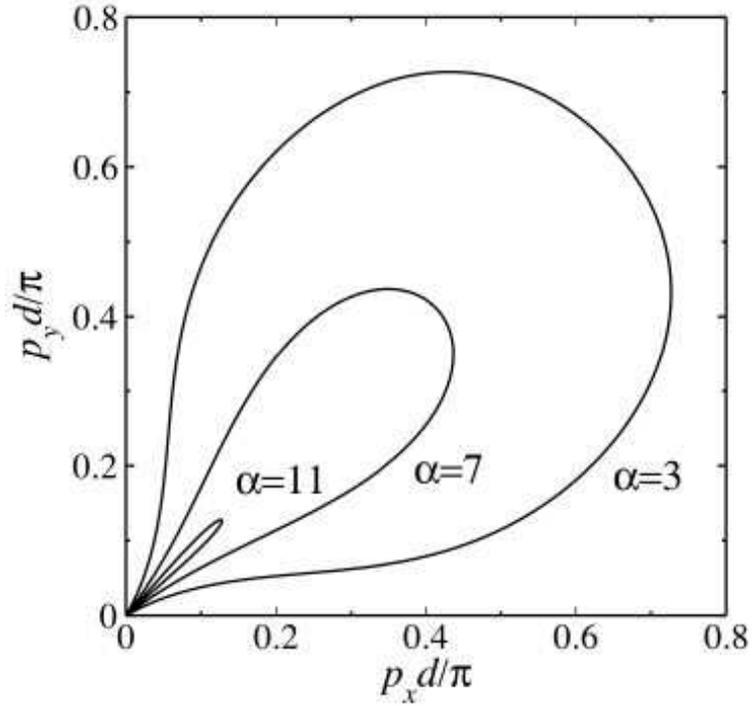}
\caption{The solution of the energy conservation condition $E_{\mathbf q}+E_{\mathbf p}=E_{\mathbf q +\mathbf p}$ in a 2D optical lattice, for $q_x=q_y=0.1\pi/d$.}
\label{qx=qy}
\end{figure}

For a long wavelength phonon $\mathbf q$ with $q_x=q_y>0$, we solve the energy conservation condition numerically. The solution is shown in Fig. \ref{qx=qy}.
There is no solution when $\alpha> 12$ as expected.
Figures \ref{energyconserv2D_qy=0} and \ref{qx=qy} clearly show that the threshold value of $\alpha$ for the disappearance of damping strongly depends on the direction of $\mathbf q$ due to the anisotropy of 2D square lattice. 
This result also holds for 3D square lattice. 

As for the 1D case, $\Gamma^\mathrm L_\mathbf q$ becomes larger than $E_{\mathbf q}$ around the threshold value of $\alpha$ in 2D and 3D optical lattices. In this case, we cannot use the simple Golden Rule expression of the Landau damping given by Eq. (\ref{Landaudamp}). We have to extend it using higher order perturbation theory in order to calculate the Landau damping when $\alpha$ is close to the threshold value \cite{Leggett}.

\subsection{Beliaev damping in a 1D optical lattice}
\label{Beliaev_damping}

We briefly discuss the Beliaev damping of the Bloch-Bogoliubov excitation in this subsection. The Beliaev damping is due to spontaneous decay of an excitation into two excitations, and thus we need to satify the energy conservation condition $E_\mathbf q=E_{\mathbf q-\mathbf p}+E_\mathbf p$. 
We focus on a 1D case in the following.

\begin{figure}[htbp]
\begin{center}
\includegraphics[width=7.5cm]{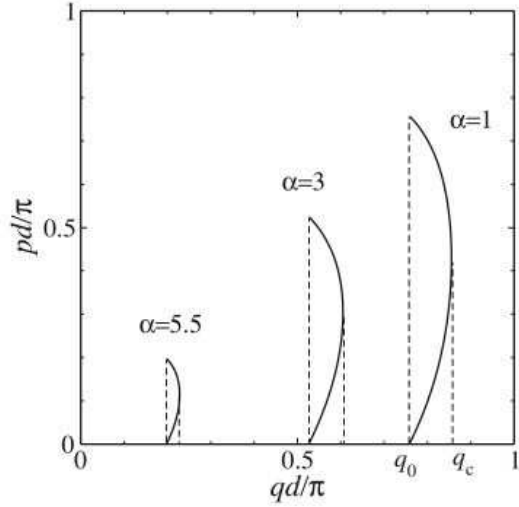}
\end{center}
\caption{The solution of the energy conservation condition $E_q=E_p+E_{q-p}$ for Beliaev damping in a 1D optical lattice.}
\label{Beliaev_energy}
\end{figure}

In Fig. \ref{Beliaev_energy}, the solution of the energy conservation condition for 1D optical lattice $E_q=E_{q-p}+E_p$ is shown in a $(q,p)$ plane. 
As predicted in Sec. \ref{1D}, one finds that the curve of the solution shrinks as $\alpha$ increases and vanishes when $\alpha>6$, which indicates the disappearance of the Beliaev damping for $\alpha>6$.

For a fixed value of $\alpha<6$, Beliaev damping is only possible when $q$ is between the threshold momenta $q_0$ and $q_{\mathrm c}$ shown in Fig. (\ref{Beliaev_energy}).
At the threshold momenta $q_0$ and $q_{\mathrm c}$, two excitations $E_q$ and $E_{q-p}$ created by the decay of an excitation $E_q$ have the same velocity. This effect was first predicted by Pitaevskii in 1959 for the phonon excitation in superfluid $^4\mathrm{He}$ \cite{Pitaevskii}. 
The two created excitations have the same quasi-momentum $q_\mathrm c/2$ and energy $E_{q_\mathrm c/2}$ at $q=q_\mathrm c$.
At $q=q_0$, one of the generated excitations is a phonon having the sound velocity $c$. Therefore, the other one also has the velocity equal to the sound velocity $c$. 

In addition to Landau damping and Beliaev damping, one also has intercollisional damping arising from two body collisions which transfer atoms between the condensate and thermal cloud at finite temperatures \cite{WG,ZNG}. Such processes also involve the energy conservation condition for three-excitation processes. Thus, the intercollisional damping also disappears when $\alpha\ge 6D$ in a $D$-dimensional optical lattice. 

\section{\label{conclusion}conclusion}

In conclusion, we have given a detailed treatment of the damping of Bogoliubov excitations associated with Bose condensates in an optical lattice at finite temperature using the tight-binding Bose-Hubbard model, extending the work we reported in Ref. \cite{TsuchiyaGriffin}.

We have used the Popov approximation in the Bose-Hubbard model to extend the usual theory to finite temperatures. As a by-product, we have calculated the number of condensate atoms per lattice site as a function of both the temperature and the lattice well depth $s$. These results may be of general interest.

We calculated the damping of Bloch-Bogoliubov excitations in 1D, 2D and 3D optical lattices. All previous work \cite{Smerzi3,Wu2} on this problem only considered dynamical instabilities. These studies did not include the dissipation processes we have considered.
We find that the Bogoliubov-Popov excitation spectrum $E_\mathbf q$ must exhibit ``anomalous dispersion" for damping processes to occur. This is analogous to the case of phonon damping in superfluid $^4\mathrm{He}$.
In the absense of this ``bending-up'' of the low $\mathbf q$ spectrum, energy conservation cannot be satisfied.
As a consequence, excitation damping is absent when $\alpha=Un^{\mathrm c 0}/J>6D$ where $D$ is the dimension of the optical lattice.

 The first studies \cite{Inguscio2} of damping of excitations were limited to 1D optical lattices along the axis of a cigar-shape magnetic trap, but one would need a much tighter magnetic trap (in the radial direction) for our 1D model results to apply.
In the more recent experiments by St\"oferle {\it et. al.} \cite{Esslinger2}, 
a 3D optical lattice is prepared first, and the lattice potential depths in two lattice axes are then made much larger than the third one to produce 2D array of tightly bound 1D optical lattices.
An analogous 2D optical lattice can be formed by choosing a much larger lattice potential depth along one lattice axis than that in other two axes.
This effectively 2D optical lattice might be better for checking our theoretical predictions than the 2D poriodic array of 1D tubes used in Ref. \cite{Esslinger1}, since excitations along the direction perpendicular to the 2D lattice potential can be neglected.  

Due to our use of a tight-binding approximation, our results are not directly applicable to the damping of excitations found in very weak optical lattices ($s<1$) \cite{Porto}.
Extension of our calculations to such weak optical lattices would be clearly of interest \cite{thesis}.

\begin{acknowledgments}
S.T. would like to thank M. Smith, E. Taylor, and D. Luxat for discussions. S.T. was supported by the Japan Society for Promotion of Science (JSPS). A.G. was supported by funds from NSERC of Canada.
\end{acknowledgments}

\appendix





\end{document}